\documentclass[iop]{emulateapj}

\usepackage{natbib}
\usepackage{booktabs}
\usepackage{url,threeparttable}
\usepackage{amssymb,amsmath}
\usepackage{txfonts,graphicx}
\catcode`\@=11
\def\gapprox{\mathrel{\mathpalette\@versim>}}
\def\lapprox{\mathrel{\mathpalette\@versim<}}
\def\@versim#1#2{\lower2.45pt\vbox{\baselineskip0pt\lineskip0.9pt
      \ialign{$\m@th#1\hfil##\hfil$\crcr#2\crcr\sim\crcr}}}
\catcode`\@=12
 
\newcommand{\hi}{H\,{\sc i}}
\newcommand{\mum}{\ensuremath{\,\mu\mbox{m}}}

\begin{document}

\title{Dust continuum emission as a tracer of gas mass in galaxies}

\author{Brent A. Groves\altaffilmark{1}\altaffiltext{1}{brent@mpia.de; Max Planck Institute for Astronomy, K\"{o}nigstuhl 17, D-69117 Heidelberg, Germany}, 
Eva Schinnerer\altaffilmark{1},  
Adam Leroy\altaffilmark{2}\altaffiltext{2}{National Radio Astronomy Observatory, 520 Edgemont Road, Charlottesville, VA 22903, USA}, 
Maud Galametz\altaffilmark{3}\altaffiltext{3}{ESO, Karl-Schwarzschild Str. 2, D-85748 Garching, Germany}, 
Fabian Walter\altaffilmark{1}, 
Alberto Bolatto\altaffilmark{4}\altaffiltext{4}{Department of Astronomy, University of Maryland, College Park, MD USA}, 
Leslie Hunt\altaffilmark{5}\altaffiltext{5}{INAF - Osservatorio Astrofisico di Arcetri, Largo E. Fermi 5, 50125 Firenze, Italy}, 
Daniel Dale\altaffilmark{6}\altaffiltext{6}{Department of Physics and Astronomy, University of Wyoming, Laramie, WY 82071 USA}, 
Daniela Calzetti\altaffilmark{7}\altaffiltext{7}{Department of Astronomy, University of Massachusetts, Amherst, MA 01003, USA}, 
Kevin Croxall\altaffilmark{8}\altaffiltext{8}{Department of Astronomy, The Ohio State University, 4051 McPherson Laboratory, 140 West 18th Avenue, Columbus, OH 43210, USA}
Robert Kennicutt Jr.\altaffilmark{9}\altaffiltext{9}{Institute of Astronomy, University of Cambridge, Madingley Road, Cambridge CB3 0HA, UK}}

\begin{abstract}
We use a sample of 36 galaxies from the KINGFISH (Herschel IR), HERACLES (IRAM CO), and THINGS (VLA HI) surveys to study empirical relations between Herschel infrared (IR) luminosities and the total mass of the interstellar gas (H$_2$ + HI). Such a comparison provides a simple empirical relationship without introducing the uncertainty of dust model fitting. We find tight correlations, and provide fits to these relations, between Herschel luminosities and the total gas mass integrated over entire galaxies, with the tightest, almost linear, correlation found for the longest wavelength data (SPIRE500). However, we find that accounting for the gas-phase metallicity (affecting the dust-to-gas ratio)  is crucial when applying these relations to low-mass, and presumably high-redshift, galaxies. The molecular (H$_2$) gas mass is found to be better correlated with the peak of the IR emission (e.g.~PACS160), driven mostly by the correlation of stellar mass and mean dust temperature. When examining these relations as a function of galactocentric radius, we find the same correlations, albeit with a larger scatter, up to radius of $r \sim 0.7 r_{25}$ (containing most of a galaxy’s baryonic mass). However, beyond that radius, the same correlations no longer hold, with increasing gas (predominantly HI) mass relative to the infrared emission. The tight relations found for the bulk of the galaxy’s baryonic content suggest that total gas masses of disk-like (non-merging/ULIRG) galaxies can be inferred from far-infrared continuum measurements in situations where only the latter are available, e.g. in ALMA continuum observations of high-redshift galaxies.
\end{abstract}

\section{Introduction}

Two of the fundamental parameters driving the evolution of galaxies across cosmic time is their total gas mass, and gas mass surface density. The total gas mass of galaxies limits the total amount of stars that can form in a galaxy at any time, while its surface density is directly linked with the rate of star formation via the observed Kennicutt-Schmidt relation \citep[e.g.][]{Kennicutt98, Leroy13}. The increase of the cosmic star formation rate (SFR) density up to redshifts 3-4 \citep[e.g.][]{Karim11} is theorized to be caused by the increase in the cosmic gas mass density (and therefore the average total gas mass of galaxies) with increasing redshift \citep[as suggested by e.g.][]{Obreschkow09}. However, directly measuring the gas mass density out to these redshifts is a difficult task.

Existing large area surveys of the  21\,cm \hi\ fine-structure line that trace the atomic gas have only measured the local Universe ($z\lapprox 0.05$, e.g.~HIPASS, \citet{Barnes01}, and ALFALFA, \citet{Haynes11}). In the near future, deeper surveys with the precursors for the Square Kilometer Array (SKA) such as DINGO \citep{Meyer09} on the Australian SKA Pathfinder (ASKAP) will reach at most $z\sim 0.4$ \citep{Duffy12}. However, it will not be until SKA is available that \hi\ will be observed in emission at $z > 1$, and even then it is expected to reach at most $z\sim1.5$ \citep{Abdalla05}. This leaves only \hi\ absorption line studies, such as damped Ly$\alpha$ systems \citep[e.g.][]{Prochaska05}, as the only determination of the evolution of the HI mass function over cosmic time.

For molecular gas, surveys of large numbers of galaxies have been limited due to the difficulty of observing the transitions from the CO molecule in the (sub--)mm. The two largest efforts have been the FCRAO Extragalactic CO Survey \citep{Young95}, which measured the CO$(1 - 0)$ line in 300 nearby galaxies, and the COLD GASS survey \citep{Saintonage11a,Saintonage11b} which measured fluxes in the CO$(1-0)$ line for a purely mass-selected sample of $\sim 350$ galaxies at $0.025 < z < 0.05$, matched in GALEX, Arecibo, and SDSS imaging. The FCRAO sample has recently been extended with CO$(1-0)$ observations of a further 59 nearby galaxies as part of the Herschel Reference Survey \citep{Boselli14a,Boselli14b}. At higher redshift the CO lines are still observable, but samples are smaller in number. While molecular gas measures were originally limited to very luminous, rare objects \citep[e.g.][]{Solomon05}, more typical galaxies are now being observed out to $z\sim0.5$ \citep[e.g., EGNoG;][]{Bauermeister13}, and farther out to $z\sim 2 - 3$ \citep[e.g., PHIBSS;][]{Tacconi13}, with a full review of these surveys available in \citet{Carilli13}. 
With ALMA now operating, the samples of high-redshift galaxy molecular gas masses will only increase. However, given its small instantaneous field-of-view, ALMA is a relatively slow survey instrument, and, while already providing impressive targeted observations, will not be suitable for very wide area blind CO surveys at $z\sim 2 - 3$ \citep{Obreschkow09}, though smaller blind searches have already been carried out in smaller regions \citep[e.g.,][]{Decarli14, Walter14}. 

Another potential tracer of the total gas mass is the infrared emission arising from dust. The amount of dust has long been known to be associated with gas column within the Milky Way \citep[e.g.][]{Jenkins74}, and dust extinction has been used to map gas columns at higher dynamic ranges and spatial resolution than available from gas emission lines \citep[e.g.][]{Lada94,Kainulainen11}. However, in extragalactic studies the lower physical resolution mixes lines of sight, making internal extinction a poor tracer of the gas column (e.g.~\citet{Kreckel13}, and see also \citet{Boquien13}), and relying on the rare occultation of relatively bright background sources is not feasible for large extragalactic studies. However, the dust continuum emission in the infrared -- submillimeter range is a direct tracer of the total dust opacity and, if a mean grain emissivity is assumed, the dust column. 

The idea of using the sub-mm continuum emission to measure the mass of the ISM has existed for a while, having been suggested by \citet{Hildebrand83}, and used by \citet{Guelin93,Guelin95} in nearby galaxies. As an independent tracer of the gas mass, the dust continuum has also been directly compared with CO observations in nearby galaxies to constrain the conversion factor between CO and gas mass (i.e.~the ``X'' factor, \citet{Israel97}, \citet{Boselli02}, \citet{Magrini11}, \citet{Sandstrom13}, and the review by \citet{Bolatto13}). Direct comparisons of the sub-mm continuum and gas content in galaxies have also been explored in \citet{Corbelli12}, \citet{Eales12}, and \citet{Scoville14}, and this correlation has already been used to estimate the total gas masses in higher redshift galaxies by \citet{Magdis12} and \citet{Scoville14}. 

Yet the dust continuum emission is not a simple direct tracer of the gas mass. The dust emission is dependent upon the temperature of the dust grains, the conversion to dust column depends upon knowledge of the grain emissivity, and finally, the conversion of dust column to gas mass depends upon the dust to gas ratio. Modelling the IR spectral energy distribution typically involves the use of both physically-based models for the dust grains and reasonable assumptions for the heating of the grains to determine the total dust column directly from the IR emission \citep[see e.g.][]{Draine07a}. This approach has obtained reasonable results in many nearby galaxies \citep[e.g.][]{Draine07b, Aniano12,Compiegne11}.  However, systematic uncertainties lie in both the dust models and the assumptions on the heating radiation field (and the parameters in each), as well as the fitting procedure itself, introducing another level of complexity in determining the interstellar gas mass. The recent work of \citet{Eales12} demonstrated that using a simple modified-blackbody model for the dust emission and a constant conversion factor from dust to gas are not unreasonable assumptions for the determination of total gas mass. They found a good correlation of the Herschel-based dust masses with total gas masses determined from HI and CO data in ten nearby galaxies, with an estimated 25\% error on the dust method. However, systematic uncertainties were found to dominate this method as well, with \citet{Eales12} demonstrating the sensitivity of the determined dust mass to the fitted dust temperature, and the sensitivity of this dust temperature to both the assumed emissivity \citep[as demonstrated in detail in ][]{Shetty09} and the observed wavelengths fitted.

We can bypass these systematic issues by comparing directly the far-IR to submm continuum emission to the various tracers of the gas phases directly. We can then determine calibrations linking the monochromatic IR emission and gas tracers by exploring a wide range of objects, which also enable the exploration of dependencies on other physical parameters like gas-phase metallicity.
A direct comparison of the sub-mm emission with gas tracers has already been explored somewhat by \citet{Bourne13}, who used a sample of 20 galaxies within the \emph{Herschel}-ATLAS survey with existing \hi\ and new CO data. They found that the sub-millimeter fluxes appear to be strongly associated with the diffuse atomic and molecular gas phases, and that the FIR/CO luminosity ratio decreases with increasing luminosity. However, even though several IR fluxes were explored, the range in the sub-mm luminosities of the galaxies was small. 
The most detailed examination to date has been by \citet{Scoville14}, who used both local luminous IR galaxies and higher redshift submillimeter galaxies with 850\mum\ fluxes to calibrate the ratio of sub-millimeter luminosity to the ISM mass assuming a constant emissivity to mass ratio. They found that the ISM mass and 850\mum\ luminosity correlated within a factor of two for their galaxy sample, and then used this correlation to determine the ISM masses in stacked submillimeter ALMA observations of high redshift galaxies. However their analysis was focussed on more luminous sources and may not hold for all galaxies.

To improve upon these surveys we use the most comprehensive nearby galaxy surveys available: the KINGFISH \citep{Kennicutt11}, THINGS \citep{Walter08}, and HERACLES \citep{Leroy09} surveys for the IR, HI, and CO data, respectively.  The galaxies in these surveys cover a wide range in metallicity, stellar mass, and star formation rates, and with all these parameters measured in a consistent manner from the associated ancillary datasets. It thus represents one of the largest homogenous datasets for CO and \hi\ in nearby galaxies, and covers the full IR spectral energy distribution (SED) from 70 to 500\mum.  In addition, the galaxies in the sample are near enough that they can be resolved and radial trends can be examined as well. This enables us to determine the correlation of sub-mm emission with gas mass over a wide range of galaxy parameters, and explore the limiting conditions where the sub-mm emission can be used.
In the following section we introduce the surveys and the galaxy sample, followed by a description of the theory in section \ref{sec:IRgaseq}, and explore a direct comparison of the integrated IR emission to the total gas masses of galaxies in section \ref{sec:IRgascorr}. We explore the issues of internal variations and radial trends in section \ref{sec:radial}, and summarize our findings in section \ref{sec:summary}.

\section{Datasets}
\subsection{Galaxy Sample}\label{sec:galaxy}

Our galaxy sample arises from the intersection of three complementary large programs surveying nearby galaxies; KINGFISH\citep{Kennicutt11}, THINGS \citep{Walter08}, and HERACLES \citep{Leroy09}.  These programs provide us with tracers of both the  integrated gas and dust masses of galaxies, and resolved maps down to $\sim$kpc scales extending well beyond the optical radii for the HI and IR data. The matched datasets result in 36 galaxies, ranging from dwarf galaxies to massive spirals, with the names and galaxy properties listed in Table \ref{tab:galaxies}. All values were taken from \citet{Kennicutt11} and references therein except for the metallicity which is from \citet{Moustakas10} when available. The sample is dominated by spiral galaxies and irregulars. The stellar masses and star formation rates both cover approximately 4 orders of magnitude, and approximately 2 orders of magnitude in specific star formation rates (sSFR$={\rm SFR}/M_{\star}$). 


\begin{deluxetable*}{lccccccccc}
\tablecaption{Galaxy Properties of the matched KINGFISH-HERACLES-THINGS sample\tablenotemark{a}.\label{tab:galaxies}}
\tablehead{\colhead{Galaxy} & \colhead{R.A.} & \colhead{Dec} & \colhead{Distance} &\colhead{P.A.} & \colhead{Major}\tablenotemark{b} & \colhead{Minor} &  \colhead{Metallicity\tablenotemark{c}} & \colhead{log M$_{\star}$} & \colhead{SFR} \\
& \colhead{(J2000)} & \colhead{(J2000)} & \colhead{(Mpc)} & \colhead{(deg)} & \colhead{(arcmin)} & \colhead{(arcmin)} & \colhead{$12+$log(O/H)} & \colhead{M$_{\odot}$}  & \colhead{M$_{\odot}$\,yr$^{-1}$}}
\startdata
DDO053 & 08:34:07 & $+$66:10:54 & 3.61 & 132 & 1.5 & 1.3 & 7.98 & 6.35 & 0.0060 \\
DDO154 & 12:54:05 & $+$27:08:59 & 4.3 & 50 & 3.0 & 2.2 & 8.02 & 6.63 & 0.0020 \\
HoI & 09:40:32 & $+$71:10:56 & 3.9 & 0 & 3.6 & 3.0 & 8.04 & 6.87 & 0.0040 \\
HoII & 08:19:05 & $+$70:43:12 & 3.05 & 15 & 7.9 & 6.3 & 8.13 & 7.59 & 0.036 \\
IC2574 & 10:28:23 & $+$68:24:44 & 3.79 & 56 & 13.2 & 5.4 & 8.23 & 8.2 & 0.057 \\
M81dwB & 10:05:31 & $+$70:21:52 & 3.6 & 140 & 0.9 & 0.6 & 8.2 & 6.36 & 0.0010 \\
NGC337 & 00:59:50 & $-$07:34:41 & 19.3 & 130 & 2.9 & 1.8 & 8.84 & 9.32 & 1.3 \\
NGC628 & 01:36:42 & $+$15:47:00 & 7.2 & 25 & 10.5 & 9.5 & 8.88 & 9.56 & 0.68 \\
NGC925 & 02:27:17 & $+$33:34:45 & 9.12 & 107 & 10.5 & 5.9 & 8.73 & 9.49 & 0.54 \\
NGC2146 & 06:18:38 & $+$78:21:25 & 17.2 & 123 & 6.0 & 3.4 & $\cdot$ & 10.3 & 7.94 \\
NGC2798 & 09:17:23 & $+$41:59:59 & 25.8 & 160 & 2.6 & 1.0 & 9.04 & 10.04 & 3.38 \\
NGC2841 & 09:22:03 & $+$50:58:35 & 14.1 & 153 & 8.1 & 3.5 & 9.19 & 10.17 & 2.45 \\
NGC2976 & 09:47:15 & $+$67:54:59 & 3.55 & 155 & 5.9 & 2.7 & 8.98 & 8.96 & 0.082 \\
NGC3077 & 10:03:19 & $+$68:44:02 & 3.83 & 45 & 5.4 & 4.5 & $\cdot$ & 9.34 & 0.094 \\
NGC3184 & 10:18:17 & $+$41:25:28 & 11.7 & 135 & 7.4 & 6.9 & 9.07 & 9.5 & 0.66 \\
NGC3198 & 10:19:55 & $+$45:32:59 & 14.1 & 35 & 8.5 & 3.3 & 8.78 & 9.83 & 1.01 \\
NGC3351 & 10:43:58 & $+$11:42:14 & 9.33 & 12 & 7.4 & 5.0 & 9.21 & 10.24 & 0.58 \\
NGC3521 & 11:05:49 & $-$00:02:09 & 11.2 & 160 & 11.0 & 5.1 & 9.06 & 10.69 & 1.95 \\
NGC3627 & 11:20:15 & $+$12:59:30 & 9.38 & 173 & 9.1 & 4.2 & 8.99 & 10.49 & 1.7 \\
NGC3938 & 11:52:49 & $+$44:07:15 & 17.9 & 28 & 5.4 & 4.9 & 9.06 & 9.46 & 1.77 \\
NGC4236 & 12:16:42 & $+$69:27:45 & 4.45 & 162 & 21.9 & 7.2 & 8.74 & 8.36 & 0.13 \\
NGC4254 & 12:18:50 & $+$14:24:59 & 14.4 & 23 & 5.4 & 4.7 & 9.08 & 9.56 & 3.92 \\
NGC4321 & 12:22:55 & $+$15:49:21 & 14.3 & 30 & 7.4 & 6.3 & 9.12 & 10.3 & 2.61 \\
NGC4536 & 12:34:27 & $+$02:11:17 & 14.5 & 130 & 7.6 & 3.2 & 9.0 & 9.44 & 2.17 \\
NGC4569 & 12:36:50 & $+$13:09:46 & 9.86 & 23 & 9.5 & 4.4 & 9.26 & 10.0 & 0.29 \\
NGC4579 & 12:37:44 & $+$11:49:05 & 16.4 & 95 & 5.9 & 4.7 & 9.22 & 10.02 & 1.1 \\
NGC4625 & 12:41:53 & $+$41:16:26 & 9.3 & 27 & 2.2 & 1.9 & 9.05 & 8.72 & 0.052 \\
NGC4631 & 12:42:08 & $+$32:32:29 & 7.62 & 86 & 15.5 & 2.7 & 8.75 & 9.76 & 1.7 \\
NGC4725 & 12:50:27 & $+$25:30:03 & 11.9 & 35 & 10.7 & 7.6 & 9.1 & 10.52 & 0.44 \\
NGC4736 & 12:50:53 & $+$41:07:14 & 4.66 & 116 & 11.2 & 9.1 & 9.04 & 10.34 & 0.38 \\
NGC5055 & 13:15:49 & $+$42:01:45 & 7.94 & 102 & 12.6 & 7.2 & 9.11 & 10.55 & 1.04 \\
NGC5457 & 14:03:13 & $+$54:20:57 & 6.7 & 39 & 28.8 & 26.0 & $\cdot$ & 9.98 & 2.33 \\
NGC5474 & 14:05:02 & $+$53:39:44 & 6.8 & 97 & 4.8 & 4.3 & 8.83 & 8.7 & 0.091 \\
NGC5713 & 14:40:12 & $-$00:17:20 & 21.4 & 11 & 2.8 & 2.5 & 9.03 & 10.07 & 2.52 \\
NGC6946 & 20:34:52 & $+$60:09:14 & 6.8 & 63 & 11.5 & 9.8 & 8.99 & 9.96 & 7.12 \\
NGC7331 & 22:37:04 & $+$34:24:56 & 14.5 & 168 & 10.5 & 3.7 & 9.05 & 10.56 & 2.74 \\
\enddata
\tablenotetext{a}{Values taken directly from \citet{Kennicutt11} and the NASA Extragalactic Database.}
\tablenotetext{b}{$R_{25}$ is equal to half the major axis.} 
\tablenotetext{c}{Metallicity shown here is the average galaxy value from \citet{Moustakas10} or \citet{Kennicutt11} when unavailable, using the \citet{Kobulnicky04} calibration.}
\end{deluxetable*}

\subsection{Infrared luminosities}
The infrared data we use in this work are all from the KINGFISH sample \citep[Key Insights into Nearby Galaxies; a Far Infrared Survey with Herschel;][]{Kennicutt11}, an imaging and spectroscopic survey of 61 nearby ($d < 30$\,Mpc) galaxies with the \emph{Herschel Space Observatory} \citep{Pilbratt10}. This survey provides imaging in 6 IR bands; 70, 100, and 160\mum\ with PACS \citep{Poglitsch10}, and 250, 350, and 500\mum\ with SPIRE \citep{Griffin10}, with the observing strategy described in detail in \citet{Kennicutt11}. All bands (PACS and SPIRE) are processed from level 1 using {\sc scanamorphos} v17.0 \citep{Roussel12}. The PACS data are calibrated using Flight Model 6, however the 160\mum\ band has a uniform correction factor of 0.925 applied to it to correct for the PACS distortion flatfield not included in the original scanamorphos reduction\footnote{see \url{http://www2.iap.fr/users/roussel/herschel/} for details.}. As the effect is of the order of 1\% for the other PACS bands, well below the photometric uncertainties, no correction is applied to these bands. 
The SPIRE bands assume factors of 97.7, 55.0, and 26.0 to convert from Jy beam$^{-1}$ to MJy sr$^{-1}$ for the SPIRE 250\mum, 350\mum, and 500\mum\ bands, respectively. 

\begin{deluxetable*}{lcccccc}
\tablecaption{Integrated IR fluxes of the 36 KINGFISH galaxies in the 6 Herschel bands.\tablenotemark{a} \label{tab:fluxes}}
\small
\tablehead{\colhead{Galaxy} &  \colhead{PACS70} &   \colhead{PACS100} &  \colhead{PACS160} &   \colhead{SPIRE250} & \colhead{SPIRE350} & \colhead{SPIRE500} \\
& \colhead{(Jy)} & \colhead{(Jy)} & \colhead{(Jy)} & \colhead{(Jy)} & \colhead{(Jy)} & \colhead{(Jy)}}
\startdata
 DDO053 & $    0.34 \pm  0.08$  & $    0.50 \pm  0.10$  & $    0.31 \pm  0.06$  & $    0.22 \pm  0.03$  & $    0.14 \pm  0.01$  & $    0.05 \pm  0.01$  \\ 
 DDO154 & $   0.00 \pm  0.07$  & $    0.49 \pm  0.07$  & $    0.12 \pm  0.08$  & $    0.18 \pm  0.06$  & $    0.12 \pm  0.03$  & $    0.05 \pm  0.02$  \\ 
    HoI & $    0.71 \pm  0.23$  & $    0.53 \pm  0.37$  & $    0.26 \pm  0.18$  & $    0.44 \pm  0.10$  & $    0.27 \pm  0.02$  & $    0.14 \pm  0.02$  \\ 
   HoII & $    3.60 \pm  0.67$  & $    3.53 \pm  0.82$  & $    3.26 \pm  0.54$  & $    1.66 \pm  0.33$  & $    0.99 \pm  0.22$  & $    0.51 \pm  0.14$  \\ 
 IC2574 & $    5.56 \pm  1.19$  & $    7.84 \pm  0.80$  & $    8.90 \pm  1.48$  & $    6.05 \pm  0.75$  & $    4.34 \pm  0.35$  & $    1.83 \pm  0.19$  \\ 
 M81dwB & $    0.09 \pm  0.05$  & $    0.17 \pm  0.11$  & $    0.11 \pm  0.06$  & $    0.17 \pm  0.01$  & $    0.10 \pm  0.01$  & $    0.04 \pm  0.01$  \\ 
NGC337 & $   13.44 \pm  0.41$  & $   20.12 \pm  0.22$  & $   18.78 \pm  0.25$  & $    9.11 \pm  0.37$  & $    4.09 \pm  0.18$  & $    1.66 \pm  0.10$  \\ 
NGC628 & $   41.07 \pm  2.48$  & $   79.11 \pm  2.60$  & $  111.20 \pm  2.49$  & $   65.36 \pm  0.93$  & $   31.55 \pm  0.50$  & $   12.46 \pm  0.24$  \\ 
NGC925 & $   12.75 \pm  1.13$  & $   25.85 \pm  0.76$  & $   35.74 \pm  0.74$  & $   26.19 \pm  0.61$  & $   14.58 \pm  0.32$  & $    6.98 \pm  0.14$  \\ 
NGC2146 & $  203.14 \pm  0.53$  & $  241.89 \pm  0.54$  & $  176.05 \pm  0.33$  & $   66.08 \pm  0.21$  & $   23.97 \pm  0.14$  & $    7.75 \pm  0.09$  \\ 
NGC2798 & $   25.14 \pm  0.24$  & $   28.65 \pm  0.35$  & $   20.53 \pm  0.25$  & $    8.10 \pm  0.03$  & $    2.95 \pm  0.03$  & $    0.92 \pm  0.02$  \\ 
NGC2841 & $   10.48 \pm  1.01$  & $   28.56 \pm  0.98$  & $   47.83 \pm  0.76$  & $   33.79 \pm  0.22$  & $   15.93 \pm  0.17$  & $    6.39 \pm  0.09$  \\ 
NGC2976 & $   20.70 \pm  1.17$  & $   37.32 \pm  1.04$  & $   44.48 \pm  1.88$  & $   25.30 \pm  1.31$  & $   11.80 \pm  0.73$  & $    4.81 \pm  0.31$  \\ 
NGC3077 & $   20.67 \pm  0.83$  & $   28.77 \pm  0.77$  & $   27.72 \pm  0.70$  & $   14.63 \pm  0.92$  & $    6.93 \pm  0.48$  & $    2.82 \pm  0.18$  \\ 
NGC3184 & $   16.40 \pm  2.11$  & $   36.60 \pm  1.08$  & $   52.34 \pm  1.00$  & $   33.03 \pm  0.33$  & $   15.31 \pm  0.25$  & $    6.22 \pm  0.13$  \\ 
NGC3198 & $   10.54 \pm  0.36$  & $   21.62 \pm  0.64$  & $   29.12 \pm  0.25$  & $   18.96 \pm  0.07$  & $    9.91 \pm  0.07$  & $    4.34 \pm  0.06$  \\ 
NGC3351 & $   27.22 \pm  1.42$  & $   48.04 \pm  1.77$  & $   52.77 \pm  0.76$  & $   33.00 \pm  0.21$  & $   14.39 \pm  0.19$  & $    5.28 \pm  0.13$  \\ 
NGC3521 & $   80.62 \pm  3.22$  & $  161.10 \pm  1.75$  & $  199.98 \pm  1.48$  & $  111.80 \pm  1.02$  & $   47.87 \pm  0.47$  & $   17.86 \pm  0.20$  \\ 
NGC3627 & $  104.75 \pm  1.41$  & $  183.48 \pm  1.08$  & $  192.74 \pm  0.93$  & $   94.25 \pm  0.63$  & $   37.39 \pm  0.27$  & $   12.95 \pm  0.17$  \\ 
NGC3938 & $   16.82 \pm  1.20$  & $   29.40 \pm  1.04$  & $   38.72 \pm  0.89$  & $   22.77 \pm  0.19$  & $   10.24 \pm  0.12$  & $    3.89 \pm  0.09$  \\ 
NGC4236 & $    8.37 \pm  1.65$  & $   13.83 \pm  1.58$  & $   18.04 \pm  1.03$  & $   11.06 \pm  0.36$  & $    7.28 \pm  0.26$  & $    3.97 \pm  0.20$  \\ 
NGC4254 & $   57.34 \pm  0.92$  & $  110.04 \pm  0.94$  & $  124.98 \pm  0.53$  & $   64.83 \pm  0.33$  & $   26.05 \pm  0.20$  & $    9.04 \pm  0.13$  \\ 
NGC4321 & $   43.33 \pm  1.23$  & $   86.57 \pm  1.09$  & $  114.04 \pm  0.66$  & $   65.87 \pm  0.32$  & $   27.70 \pm  0.26$  & $    9.70 \pm  0.15$  \\ 
NGC4536 & $   40.22 \pm  0.64$  & $   54.29 \pm  0.46$  & $   53.77 \pm  0.29$  & $   27.95 \pm  0.09$  & $   12.49 \pm  0.12$  & $    5.01 \pm  0.09$  \\ 
NGC4569 & $   15.49 \pm  0.52$  & $   31.32 \pm  0.62$  & $   39.08 \pm  0.44$  & $   22.05 \pm  0.23$  & $    9.48 \pm  0.15$  & $    3.43 \pm  0.06$  \\ 
NGC4579 & $   10.10 \pm  0.35$  & $   24.53 \pm  0.52$  & $   33.39 \pm  0.45$  & $   20.02 \pm  0.09$  & $    8.66 \pm  0.06$  & $    3.19 \pm  0.03$  \\ 
NGC4625 & $    1.35 \pm  0.20$  & $    3.34 \pm  0.37$  & $    4.42 \pm  0.24$  & $    2.66 \pm  0.07$  & $    1.35 \pm  0.06$  & $    0.58 \pm  0.03$  \\ 
NGC4631 & $  141.22 \pm  1.14$  & $  233.79 \pm  1.17$  & $  238.95 \pm  1.16$  & $  123.92 \pm  0.23$  & $   56.22 \pm  0.17$  & $   22.61 \pm  0.12$  \\ 
NGC4725 & $    8.49 \pm  1.00$  & $   25.47 \pm  1.59$  & $   44.69 \pm  0.56$  & $   31.64 \pm  0.28$  & $   16.58 \pm  0.20$  & $    7.10 \pm  0.12$  \\ 
NGC4736 & $  103.66 \pm  2.44$  & $  161.73 \pm  4.27$  & $  137.47 \pm  1.63$  & $   66.97 \pm  0.66$  & $   27.69 \pm  0.42$  & $   10.09 \pm  0.22$  \\ 
NGC5055 & $   74.51 \pm  3.61$  & $  174.72 \pm  3.13$  & $  235.81 \pm  2.00$  & $  145.67 \pm  0.94$  & $   63.60 \pm  0.46$  & $   24.05 \pm  0.32$  \\ 
NGC5457 & $  127.51 \pm  7.83$  & $  255.02 \pm  5.17$  & $  320.70 \pm  1.89$  & $  202.81 \pm  0.92$  & $   98.17 \pm  0.82$  & $   42.00 \pm  0.44$  \\ 
NGC5474 & $    3.23 \pm  0.82$  & $    4.88 \pm  0.68$  & $    6.91 \pm  0.43$  & $    5.14 \pm  0.18$  & $    3.02 \pm  0.09$  & $    1.48 \pm  0.06$  \\ 
NGC5713 & $   28.83 \pm  0.25$  & $   40.79 \pm  0.17$  & $   37.19 \pm  0.33$  & $   16.26 \pm  0.08$  & $    6.25 \pm  0.06$  & $    2.08 \pm  0.03$  \\ 
NGC6946 & $  250.38 \pm  5.60$  & $  447.29 \pm  6.29$  & $  512.75 \pm  4.63$  & $  268.77 \pm  3.58$  & $  110.69 \pm  2.16$  & $   39.46 \pm  0.94$  \\ 
NGC7331 & $   68.11 \pm  0.73$  & $  134.51 \pm  1.01$  & $  167.65 \pm  0.90$  & $   92.17 \pm  0.59$  & $   39.94 \pm  0.31$  & $   15.13 \pm  0.15$  \\ 
\enddata
\tablenotetext{a}{Flux uncertainties include only uncertainties in background subtraction and noise.}
\end{deluxetable*}

To measure the fluxes of each galaxy, we first convolve all bands to a beam size of $\sim 36\arcsec$, the resolution of the SPIRE 500\mum\ beam (the lowest resolution band), using the convolution kernels of \citet{Aniano12}. We then use the galaxy centres listed in Table \ref{tab:galaxies} and elliptical apertures as used by \citet{Dale12} to determine the total integrated galaxy flux. These elliptical apertures have the same position angles and axis ratios as listed in Table \ref{tab:galaxies}, and encompass the full optical and IR light, and typically extend to $\sim2$ optical radii. The background for all images was assumed to be flat  and was determined as in \citet{Dale12}, by sampling empty regions around our elliptical apertures and calculating the background mean and uncertainty, as well as the pixel noise. To determine the luminosities we assume the distances given in Table \ref{tab:galaxies}. The total fluxes and uncertainties for all Herschel bands  for each galaxy are given in Table \ref{tab:fluxes}. These fluxes differ slightly from those in \citet{Dale12} as updated beam sizes were assumed for all bands leading to different correction factors, and the convolution to the SPIRE 500\mum\ beam size also affected results (though predominantly with the resolved investigations in section \ref{sec:radial}).

In addition to the IR photometry, as the successor to the SINGS survey \citep{Kennicutt03}, a large ancillary dataset exists for the KINGFISH galaxies, including determinations of their metallicity, stellar masses and star formation rates, as shown in Table \ref{tab:galaxies}. The metallicities of our sample are determined in \citet{Moustakas10}, using the calibration of \citet{Kobulnicky04}, and cover a range of 2 dex. Different metallicity calibrations, such as \citet[][PT05]{Pilyugin05}, will introduce systematic offsets from the calibration used here (as demonstrated for several calibrations by \citet{Kewley08}, and as can be seen for the KINGFISH galaxies, specifically in Table 1 in \citet{Kennicutt11}).

\subsection{Atomic gas masses}
To determine the neutral gas mass of our galaxy sample we use the THINGS HI sample of nearby galaxies \citep{Walter08}\footnote{The THINGS data are available from \url{http://www.mpia-hd.mpg.de/THINGS}.}, with the addition of the Leroy et al. (in prep.) sample that extends the THINGS sample to a larger number of nearby galaxies. 

\begin{deluxetable}{lccc}
\tablecaption{Integrated gas masses and 500\mum\ luminosity of the 36 KINGFISH galaxies. \label{tab:masses}}
\tablehead{\colhead{Galaxy} & \colhead{L$_{500}$} & \colhead{M$_{\rm HI}$} & \colhead{M$_{\rm H2}$} \\
 & \colhead{$10^6$ L$_{\odot}$} & \colhead{$10^7$ M$_{\odot}$} & \colhead{$10^7$ M$_{\odot}$} }
\startdata
 DDO053 & $    0.12 \pm  0.01$  & $    3.88 \pm   0.58$  & $    <0.02$  \\ 
 DDO154 & $    0.19 \pm  0.01$  & $    9.20 \pm   1.38$  & $    <0.01$  \\ 
    HoI & $    0.40 \pm  0.02$  & $   10.06 \pm   1.51$  & $    <0.03$  \\ 
   HoII & $    0.89 \pm  0.07$  & $   31.60 \pm   4.74$  & $    <0.02$  \\ 
 IC2574 & $    4.93 \pm  0.09$  & $   71.46 \pm  10.72$  & $    0.82 \pm   0.08$  \\ 
 M81dwB & $    0.11 \pm  0.01$  & $    0.92 \pm   0.14$  & $    0.00$  \\ 
NGC337 & $  116.06 \pm  0.63$  & $  330.82 \pm  49.62$  & $   36.35 \pm   1.82$  \\ 
NGC628 & $  121.05 \pm  0.72$  & $  260.53 \pm  39.08$  & $  127.41 \pm   6.37$  \\ 
NGC925 & $  108.85 \pm  0.37$  & $  448.59 \pm  67.29$  & $   26.98 \pm   1.35$  \\ 
NGC2146 & $  429.99 \pm  1.24$  & $  188.07 \pm  28.21$  & $  797.18 \pm  39.86$  \\ 
NGC2798 & $  115.18 \pm  0.93$  & $   72.72 \pm  10.91$  & $  237.92 \pm  11.90$  \\ 
NGC2841 & $  237.98 \pm  0.81$  & $  450.51 \pm  67.58$  & $   91.54 \pm   4.58$  \\ 
NGC2976 & $   11.36 \pm  0.10$  & $   12.72 \pm   1.91$  & $    7.00 \pm   0.35$  \\ 
NGC3077 & $    7.76 \pm  0.05$  & $   28.94 \pm   4.34$  & $    1.70 \pm   0.09$  \\ 
NGC3184 & $  159.61 \pm  0.92$  & $  302.38 \pm  45.36$  & $  180.82 \pm   9.04$  \\ 
NGC3198 & $  161.93 \pm  0.44$  & $  506.31 \pm  75.95$  & $   62.42 \pm   3.12$  \\ 
NGC3351 & $   86.09 \pm  0.54$  & $   95.24 \pm  14.29$  & $   93.82 \pm   4.69$  \\ 
NGC3521 & $  419.94 \pm  0.87$  & $  811.43 \pm 121.71$  & $  392.12 \pm  19.61$  \\ 
NGC3627 & $  213.68 \pm  0.61$  & $   87.40 \pm  13.11$  & $  280.36 \pm  14.02$  \\ 
NGC3938 & $  233.44 \pm  1.84$  & $  432.18 \pm  64.83$  & $  232.21 \pm  11.61$  \\ 
NGC4236 & $   14.75 \pm  0.14$  & $  203.74 \pm  30.56$  & $    0.30 \pm   0.02$  \\ 
NGC4254 & $  351.59 \pm  0.90$  & $  381.34 \pm  57.20$  & $  661.71 \pm  33.09$  \\ 
NGC4321 & $  371.87 \pm  0.98$  & $  257.89 \pm  38.68$  & $  633.54 \pm  31.68$  \\ 
NGC4536 & $  197.38 \pm  0.72$  & $  363.38 \pm  54.51$  & $  173.33 \pm   8.67$  \\ 
NGC4569 & $   62.55 \pm  0.31$  & $   19.86 \pm   2.98$  & $  126.50 \pm   6.33$  \\ 
NGC4579 & $  160.90 \pm  0.60$  & $   57.18 \pm   8.58$  & $  225.79 \pm  11.29$  \\ 
NGC4625 & $    9.34 \pm  0.16$  & $   32.12 \pm   4.82$  & $    2.48 \pm   0.12$  \\ 
NGC4631 & $  246.09 \pm  0.31$  & $  616.84 \pm  92.53$  & $  150.09 \pm   7.50$  \\ 
NGC4725 & $  188.45 \pm  0.79$  & $  294.52 \pm  44.18$  & $   62.08 \pm   3.10$  \\ 
NGC4736 & $   41.08 \pm  0.24$  & $   40.70 \pm   6.10$  & $   55.26 \pm   2.76$  \\ 
NGC5055 & $  284.20 \pm  0.94$  & $  365.07 \pm  54.76$  & $  312.90 \pm  15.65$  \\ 
NGC5457 & $  353.44 \pm  0.74$  & $  888.70 \pm 133.31$  & $  230.70 \pm  11.54$  \\ 
NGC5474 & $   12.83 \pm  0.16$  & $   69.06 \pm  10.36$  & $    0.38 \pm   0.02$  \\ 
NGC5713 & $  178.74 \pm  0.79$  & $  197.49 \pm  29.62$  & $  374.65 \pm  18.73$  \\ 
NGC6946 & $  342.04 \pm  1.38$  & $  409.18 \pm  61.38$  & $  594.70 \pm  29.73$  \\ 
NGC7331 & $  596.23 \pm  1.21$  & $  901.71 \pm 135.26$  & $  487.04 \pm  24.35$  \\ 
\enddata
\end{deluxetable}

To determine the HI masses of the galaxies, we use the integrated HI moment 0 maps with robust weighting \citep[see][ for details]{Walter08}. To first convert the HI maps from their units of Jy\,beam$^{-1}\,\rm{m\,s}^{-1}$ to M$_{\odot}\, pc^{-2}$, we multiply by \citep[based on  equations 1 and 5 from][]{Walter08},
\begin{equation}\label{eqn:HIscale}
\Sigma_{\rm HI}=8.667\times10^{3}~D_{\rm Mpc}^2 S_{\rm HI}[{\rm Jy\, beam^{-1}}]/(B_{\rm maj}\times B_{\rm min}),
\end{equation}
where $B_{\rm maj}$ and $B_{\rm min}$ are the beam major and minor axes in arcseconds, respectively. 
$D_{\rm Mpc}^2$ is the galaxy distance in Mpc (from Table \ref{tab:galaxies}).
We then convolve the maps using first gaussians to circularize the beam shapes and from there to the SPIRE 500\mum\ resolution using the convolution kernels of \citet{Aniano12}. The integrated HI masses are then determined using the same apertures as for the infrared fluxes and are given in Table \ref{tab:masses} with uncertainties and SPIRE 500\mum\ luminosities.  Note that for some galaxies the HI emission extends beyond the chosen aperture that encompasses the optical and IR emission, which we discuss further on.

\subsection{Molecular gas masses} 

For the molecular gas masses of our galaxy sample we use the HERACLES CO(2-1) survey of nearby galaxies \citep{Leroy09}\footnote{The HERACLES data are available from \url{http://www.mpia-hd.mpg.de/HERACLES/Overview.html}.}, using the integrated moment 0 maps. To determine the total CO luminosity we first convolve the moment 0 maps using the convolution kernels of \citet{Aniano12}, and then integrate using the same apertures as for the infrared and HI maps. If these apertures extended further than the available moment 0 maps, we assumed the flux in these regions was negligible and therefore 0. To convert the CO luminosity to molecular gas mass we used equation 3 from \citet{Leroy09}, 
\begin{equation}
{\rm M}_{\rm H_2}[M_{\odot}]= \alpha_{\rm CO}R_{21}L_{\rm CO}[{\rm K\,km\,s^{-1} pc^{2}}].
\end{equation}
We assume a fixed conversion factor from the CO(2-1) luminosity to H$_2$ mass for all galaxies, based on the Milky Way conversion factor, $\alpha_{\rm CO(1-0)}=4.4$ M$_{\odot}\,\rm{pc^{-2}\,(K\,km\,s^{-1})^{-1}}$ \citep{Solomon87, Abdo10, Bolatto13}, and a fixed line ratio of CO(2-1) to CO(1-0) of $R_{21}=0.8$ \citep[the typical value found for the HERACLES galaxies in][though correctly this ratio is dependent upon the mean gas properties such as temperature]{Leroy09}. 
The final H$_2$ mass for each galaxy is given in Table \ref{tab:masses}.
The total gas mass for all galaxies is then taken simply as the sum of the HI and molecular gas mass within our chosen apertures. 

Given that our galaxies cover over 2 orders of magnitude in stellar mass and gas-phase metallicity, an assumption of a constant CO luminosity to molecular gas mass conversion factor may lead to biases, especially as different conversion factors have already been determined for several of the galaxies in our sample \citep[see e.g.][]{Bolatto13,Sandstrom13}. We discuss this point further in the text, but this assumption must be considered when using any of our determined relationships.

\section{The monochromatic IR---Gas Correlation} \label{sec:IRgascorr}

\subsection{Dust emission and gas mass}\label{sec:IRgaseq}

The use of the IR emission to trace the total amount of gas is premised on the simple assumption that gas and dust are always associated. Assuming that the dust has a single temperature, ${\rm T}_{\rm d}$, the IR luminosity of a galaxy at a frequency, $\nu$, is linked to the total gas mass of a galaxy M$_{\rm gas}$ via the equation,
\begin{equation}\label{eqn:dustlum}
L_{\nu}={\rm M_{gas}}\left(\frac{D}{G}\right) 4\pi B_{\nu}({\rm T}_{\rm d})\kappa_{\nu},
\end{equation}
where $(D/G)$ is the dust to gas mass ratio, $B_{\nu}$ is the Planck blackbody function, and $\kappa_{\nu}$ is the dust emissivity at the frequency $\nu$.

A broad range of IR colors is seen within and between galaxies indicating variation of the mean dust temperature \citep[see, e.g.][]{Dale12}, driven by the variation of radiation field strengths experienced by dust in galaxies.  However, we can limit the impact of these temperature distributions on the above equation in two ways; firstly, for most of the KINGFISH galaxies, it is possible to reproduce their IR SED by having the bulk of the dust heated by a single diffuse radiation field, with only a small fraction being heated by radiation fields stronger than this \citep[see, e.g.][]{Draine07b, Aniano12}, thus assuming a single temperature within galaxies for the longer wavelengths is a reasonable assumption. Secondly, by concentrating on longer wavelengths that are well past the peak of the IR SED \citep[at $\sim100 - 200$\mum, see e.g.][]{Dale12}, we move closer to the Rayleigh-Jeans tail of the black body radiation which is only linearly sensitive to temperature. \citet[][particularly their Figure 7]{Smith12} and \citet[][particularly their Figure 12]{Auld13} have both shown that the assumption of a single temperature is reasonable, finding the temperatures of single modified blackbody fits to the integrated 100-500\mum\ IR SEDs of galaxies in the Herschel Reference Survey and Herschel Virgo Cluster Survey, respectively, lying within a factor of 2 ($25\pm10$\,K and $\sim20\pm8$\,K, respectively, with Early type galaxies having higher temperatures). Such a small range of temperatures result in  only a factor of $\sim$2 uncertainty in the dust mass estimates and hence a factor of 2 in the gas mass determined from the IR in the linear regime.

The dust-to-gas mass ratio and dust emissivity at a given wavelength are somewhat degenerate, with the determination of one relying on an estimate of the other. While estimates exist for these quantities in the local group, it is uncertain how they will vary across different galaxies. The emissivity is dependent on the composition of dust \citep{Zubko04}, and there have been suggestions that the emissivity will vary in molecular environments. The emissivity has also been suggested to vary with metallicity, such as seen in the LMC by \citet{Galliano11}, however \citet{RemyRuyer13} found no strong trends when comparing more massive galaxies with dwarf galaxies. However, if it is assumed that emissivity at longer wavelengths does not change between galaxies, a clear trend of a decreasing dust-to-gas ratio with the metallicity of the galaxy is seen across the local group \citep{Leroy11} and nearby galaxies \citep{Sandstrom13}, and even if the emissivity is left free, a lower dust-to-gas ratio is seen in low metallicity dwarf galaxies \citep{RemyRuyer14} . This suggests that the dust to gas ratio must be accounted for in any empirical calibration, as we discuss below.

Although Equation \ref{eqn:dustlum} indicates the direct link of gas mass and dust emission via the association of dust and gas, IR emission will also be indirectly associated with gas mass, or more precisely gas mass surface density via the Kennicutt-Schmidt relation \citep[KS relation, see e.g.][]{Kennicutt98}. An increase in the gas mass surface density has been empirically demonstrated to correlate on average with an increase in the star formation rate surface density on kpc scales \citep[see e.g.,][]{Kennicutt98, Bigiel08, Leroy13}. Given this increase in the SFR surface density, the interstellar radiation field that heats the dust is also expected to be higher, leading in turn to higher average dust temperatures and correspondingly higher IR continuum emission. This secondary correlation may actually limit the dispersive effect that dust temperature has on the gas mass -- IR emission connection, and may even lead to a tighter correlation where the IR luminosity is more responsive to gas mass (i.e.~where an increase in gas mass leads to both an increase in dust mass and increase in dust temperature due to the increased SFR, and thus a greater increase in IR luminosity). 
However, unlike the correlation of dust mass and gas mass, the KS relation will act non-linearly on the IR emission. The shortest dust wavelengths that measure the hot dust and peak of the IR emission will see the greatest impact of the increase in dust temperature, and thus also be more sensitive to any variation in the KS relation.

\subsection{IR emission versus total gas mass}\label{sec:IRgastot}

\begin{figure*}
\includegraphics[width=\hsize]{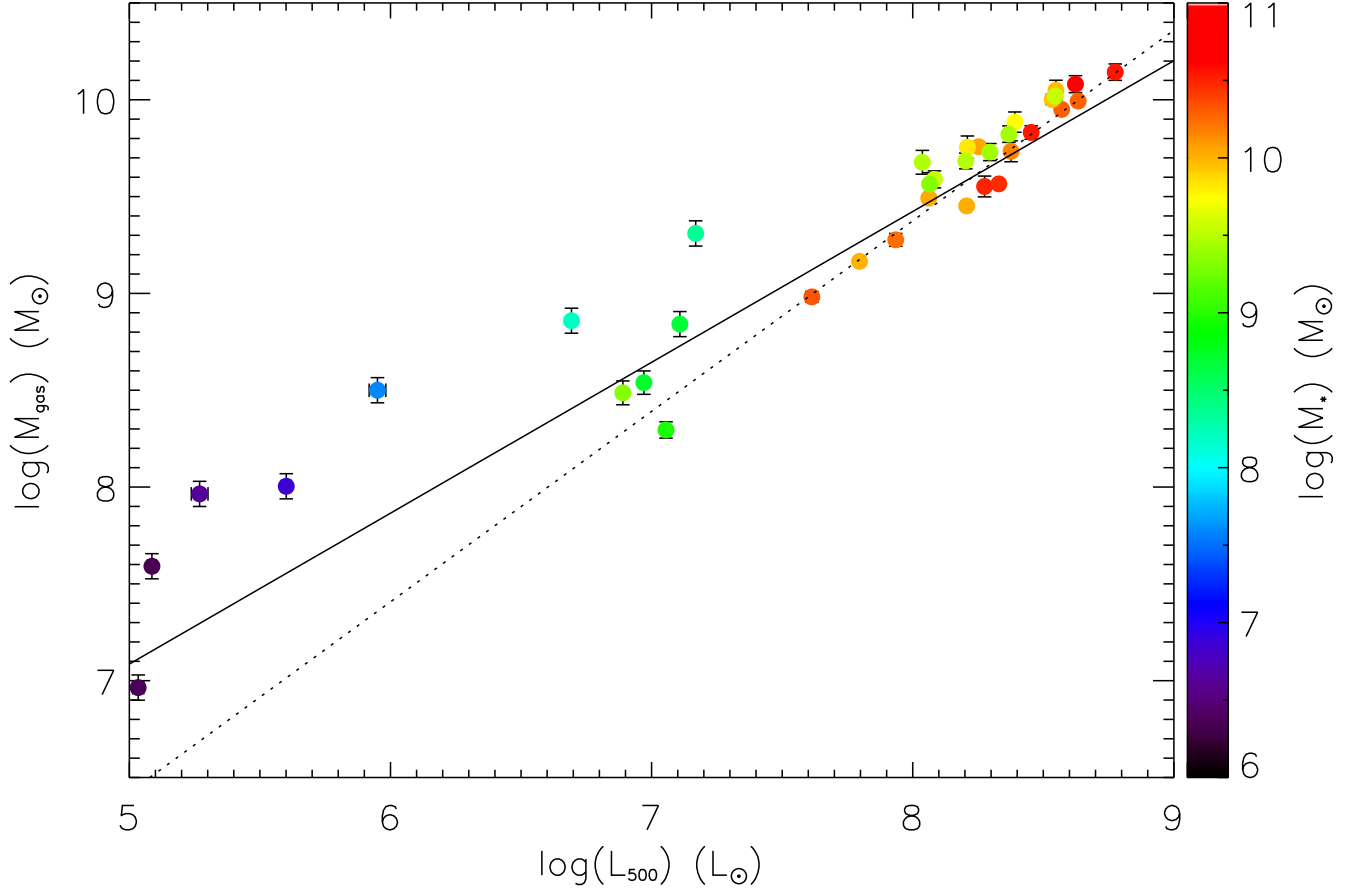}
\caption{SPIRE 500 Luminosity (L$_{500}$) versus total gas mass M$_{gas}$ for the KINGFISH sample. The colors correspond the total stellar mass of the galaxy, as indicated by the color bar to the right. Over plotted is the linear fit to the full sample (solid line), and to only the non-dwarf galaxies (log(M$_{\star}/\rm{M}_{\odot}) > 9$, dashed line), with the fit values given in Table \ref{tab:linearfitgas}. }\label{fig:500_Mgas}
\end{figure*}

As a first comparison, we determine the simple correlations between the IR luminosities and total gas mass in galaxies. In Table \ref{tab:pearson}, we show the Pearson correlation coefficient (\& Spearman rank coefficient) for all \emph{Herschel} bands against the total gas mass. 
A strong correlation is observed with all bands, with the strongest correlation existing between the longest wavelength (i.e.~SPIRE 500\mum) and the total gas mass. This correlation is clearly seen in Figure \ref{fig:500_Mgas}, where we plot the SPIRE500 luminosity ($\nu L_{\nu}$) against the total gas mass. 

\begin{table}
\caption{Correlation coefficients for $\rm log(M_{gas})$ against individual Herschel bands. }\label{tab:pearson}
\begin{tabular}{lrr}
\tableline
Band (log$(\nu{\rm L}_{\nu})$& Pearson & Spearman \\
\tableline
\tableline
    PACS70 &  0.926 &  0.821  \\
   PACS100 &  0.938 &  0.893 \\
   PACS160 &  0.943 &  0.918 \\
  SPIRE250 &  0.949 &  0.941 \\
  SPIRE350 &  0.957 &  0.950 \\
  SPIRE500 &  0.962 &  0.964 \\
\end{tabular}
\end{table}

The strong correlations seen with each band suggest that it is entirely reasonable to use broad-band IR photometry to estimate the total gas mass of galaxies from the IR emission. Based on the high Pearson correlation coefficients, we performed robust linear fits to each \emph{Herschel} band luminosity versus the total gas mass, of the form, $\log({\rm M}_{\rm gas})=A+B\times$Band, where Band is the log of the luminosity in the given band (in log$(\nu{\rm L}_{\nu})$). The coefficients $A$ and $B$ are given in Table \ref{tab:linearfitgas} for each band. The fit for the SPIRE500--M$_{\rm gas}$ relation is shown by the solid line on Figure \ref{fig:500_Mgas}. In Table \ref{tab:linearfitgas},  we also give the dispersion, $\sigma$, of the galaxies around each relation, which, as suggested by the correlation coefficients, decreases with increasing wavelength. To compare with previous comparisons of dust and gas correlations, we have also limited the sample to the more massive objects, with log(M$_{\star}/\rm{M}_{\odot}) > 9$. The fits to these galaxies show both a smaller dispersion and steeper slopes than the full sample. The SPIRE 500 fit is the tightest, with only a 30\% dispersion, similar to that found by \citet{Eales12} in their comparison of gas masses with dust masses determined from fits to the IR SED in 10 nearby galaxies.  

Given that the full extent of a galaxy in the IR may not always be available, we have also repeated the fits using smaller apertures that are truncated at an optical radius, $R_{25}$, using the values for the major and minor axes presented in Table \ref{tab:galaxies}.  These apertures are typically 30\% smaller than the IR-encompassing apertures from \citet{Dale12} that we use. However, as the IR light is generally concentrated to the optical disk, these smaller apertures contain a significant fraction ($>90$\%) of the IR light for most galaxies. Only a few galaxies with clearly extended IR emission, mostly the smaller galaxies such as NGC\,3077, have significantly reduced IR fluxes due to the smaller aperture. Similarly, most of the molecular gas is totally contained within an optical radius for all galaxies. Yet the atomic gas can extend well beyond an optical radius, and is observed at greater than $3R_{25}$ in some galaxies. Truncating at $R_{25}$ reduces the total gas mass because of this extended emission (this is discussed further in \S \ref{sec:radial}).  We have performed these fits only for the three SPIRE bands, as these show the strongest correlations in Table \ref{tab:pearson}, and are the most sensitive to extended emission (and thus show the greatest change with reduced apertures). All fits to the reduced apertures, both to the full and non-dwarf galaxy sub-sample, find slightly flatter slopes than the original apertures,  leading also to a larger intercept. This trend is driven by the dwarf galaxies,  which have relatively larger HI gas disks, such that the smaller aperture preferentially reduces the gas mass relative to the IR, flattening the overall slope. Thus biasing an aperture to measure the IR luminosity of galaxy will tend to underestimate the total gas mass, approximately by 15\% based on our galaxy sample.

\begin{table*}[t]
\begin{center}
\caption{Coefficients of linear fits of the form $\log({\rm M}_{\rm gas})=A+B\times$Band. 
}\label{tab:linearfitgas}
\begin{tabular}{lrrr|rrr}
\tableline
 & & All & & & log(M$_{\star}/{\rm M}_{\odot}) > 9$ & \\
Band (log$(\nu{\rm L}_{\nu})$& $A$ & $B$ & $\sigma$ & $A$ & $B$ & $\sigma$\\
\tableline
\tableline
   PACS70 &  4.15 &  0.55 &  0.30 &  4.6 &  0.50 &  0.27 \\
   PACS100 &  3.62 &  0.60 &  0.27 &  3.27 &  0.63 &  0.23  \\
   PACS160 &  3.52 &  0.61 &  0.26 &  1.91 &  0.78 &  0.18  \\
  SPIRE250 &  3.17 &  0.69 &  0.25 &  1.17 &  0.90 &  0.15  \\
  SPIRE350 &  3.08 &  0.74 &  0.24 &  1.17 &  0.90 &  0.14  \\
  SPIRE500 &  3.19 &  0.78 &  0.23 &  1.44 &  0.99 &  0.10  \\
\tableline
\end{tabular}
\end{center}
\end{table*}

\begin{table*}[t]
\begin{center}
\caption{Linear fit coefficients for galaxies measured within $R_{25}$. 
}\label{tab:linearfitgasR25}
\begin{tabular}{lrrr|rrr}
\tableline
 & & All & & & log(M$_{\star}/{\rm M}_{\odot}) > 9$ & \\
Band (log$(\nu{\rm L}_{\nu})$& $A$ & $B$ & $\sigma$ & $A$ & $B$ & $\sigma$\\
\tableline
\tableline
  SPIRE250 &  3.73 &  0.64 &  0.28 &  1.57 &  0.86 &  0.14 \\
  SPIRE350 &  3.63 &  0.68 &  0.25 &  1.49 &  0.92 &  0.10 \\
  SPIRE500 &  3.74 &  0.72 &  0.23 &  1.72 &  0.96 &  0.09 \\
\tableline
\end{tabular}
\end{center}
\end{table*}

We find that none of the fits are linear, neither the total gas masses and the sample limited within $R_{25}$, nor the massive galaxy sub-samples in both, with all slopes less than 1. As discussed in the previous section, three factors may be playing a role in the non-linear slopes: the systematic variation of the dust temperature, the dust-to-gas ratio, and the variation of the CO luminosity to molecular gas mass ($\alpha_{\rm CO}$).  As the slope becomes flatter at shorter wavelengths, dust temperature must play at least some role. The sub-linear trend observed in the correlation at all wavelengths suggests that the Kennicutt-Schmidt relation may also be contributing to the correlation, that is higher gas masses lead to a higher SFRs, which in turn heats the dust to higher average temperatures. 

The dependence upon the CO conversion factor is a major source of uncertainty in these fits, and remains a general uncertainty in determining the total gas masses in galaxies. Our assumed value of $\alpha_{\rm CO(1-0)}=4.4$ M$_{\odot}\,\rm{pc^{-2}\,(K\,km\,s^{-1})^{-1}}$ is the suggested value of the Milky Way conversion factor given in the extensive review of \citet{Bolatto13}. However, the analysis of $\alpha_{\rm CO}$ in some of the same galaxies as our sample by \citet{Sandstrom13} found a slightly lower value ($\alpha_{\rm CO}=3.6$), and a large scatter between, and within, the galaxies. Changing the value of $\alpha_{\rm CO}$ we assume significantly affects only the more massive spirals (M$_{\star} > 10^{9}\,{\rm M}_{\odot}$), as the dwarf galaxies are all HI dominated, as can be seen from Table \ref{tab:masses}. Increasing/decreasing $\alpha_{\rm CO}$ by 0.7 dex \citep[larger than the dispersion between galaxies found by][]{Sandstrom13} leads to a steepening/flattening of the determined linear fits. However, when only the galaxies with M$_{\star} > 10^{9}\,{\rm M}_{\odot}$ are examined, the slope remains approximately constant, with a systematic shift up and down in the relation, scaling with our scaling of $\alpha_{\rm CO}$. However, given the work of \citet{Sandstrom13} and review by \citet{Bolatto13}, there is no a priori reason why our normal spiral galaxy sample should have a different $\alpha_{\rm CO}$ to that assumed in our work.

\begin{figure*}
\includegraphics[width=\hsize]{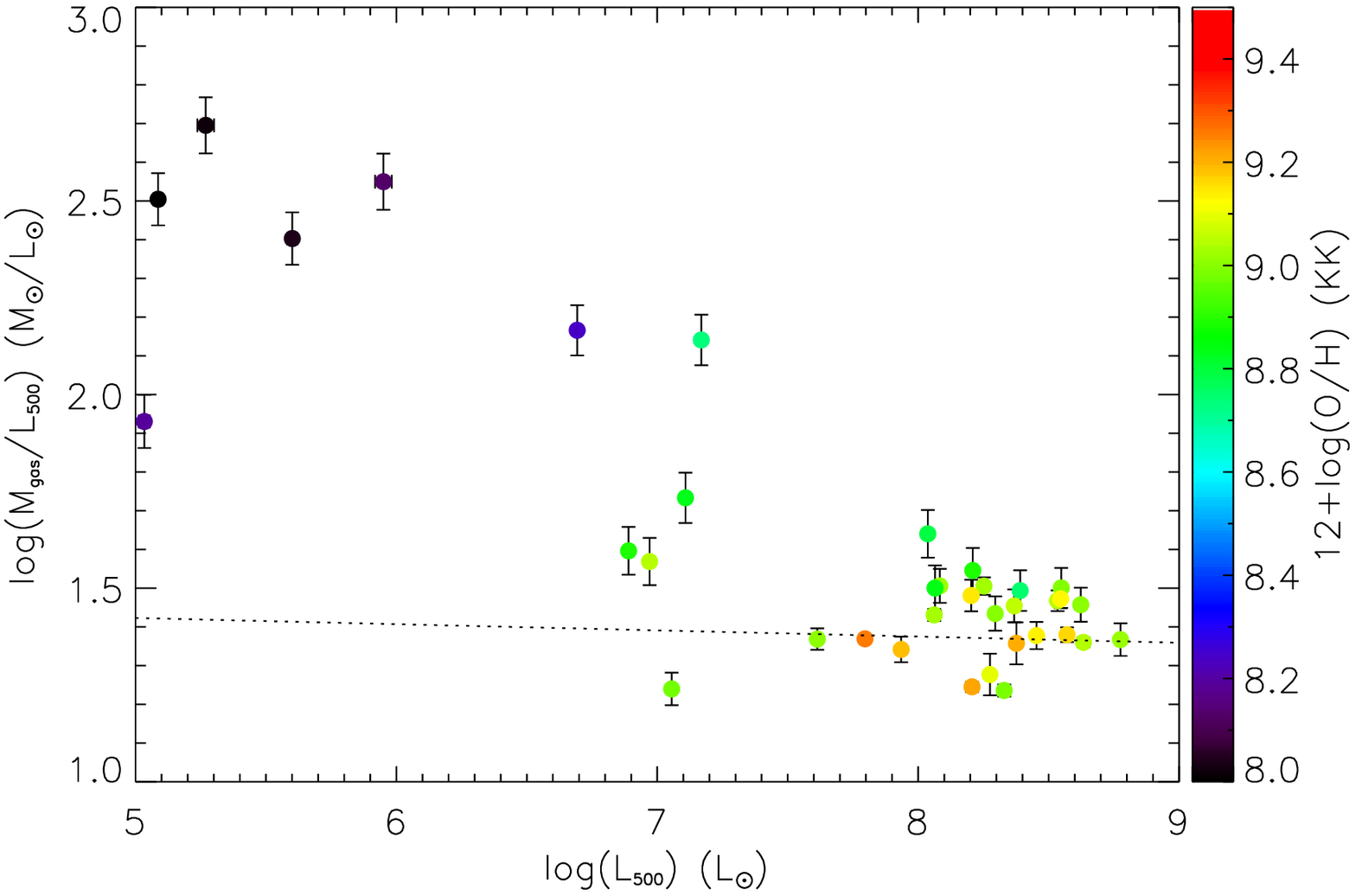}
\caption{SPIRE 500\mum\ luminosity (L$_{500}$) against the ratio of the total gas mass M$_{gas}$ to L$_{500}$. The colors in this figure show the average metallicity of the galaxies, as measured by \citet{Moustakas10}, using the \citet{Kobulnicky04} conversion. This figure clearly indicates the linear relationship between M$_{\rm gas}$ and L$_{500}$ for non-dwarf galaxies, and the offset of the low stellar mass (low metallicity) galaxies. }\label{fig:L500Mgas_L500}
\end{figure*}

However, there are two further issues with the simple linear fits revealed by Figure \ref{fig:500_Mgas}; a large part of the correlation of gas mass with IR luminosity is simply galaxy mass scaling, with galaxies with the highest stellar masses lying towards the top right, as indicated by the color of the points. Secondly, the lower mass galaxies tend to lie above the relation, leading to a much greater scatter, and limit the determination of the gas mass.  In Figure \ref{fig:L500Mgas_L500}, we remove the mass weighting intrinsic to a luminosity-luminosity plot by plotting the ratio M$_{\rm gas}/{\rm L}_{500}$ against ${\rm L}_{500}$, which also emphasizes the scatter of the low-mass galaxies. 
The color scale in this figure reveals the average metallicity of the galaxy, as measured by \citet{Moustakas10} (using the \citet{Kobulnicky04} conversion). The similarity in colors between the two plots demonstrates the mass-metallicity relation.  
In this figure, it is clear that the more massive normal galaxies show a much more linear relation, with the offset caused by the low-mass, low-metallicity galaxies in the sample. 
In Table \ref{tab:linearfitgas} we also include a fit to only the galaxies with log(M$_{\star}/{\rm M}_{\odot}) > 9)$ (non-dwarf galaxies, hereafter ``normal'' galaxies), meaning a $12+\log({\rm O/H})_{\rm KK} \gapprox 8.6$ using the \citet{Kewley08} mass-metallicity relation). This is shown as the dashed line in Figures \ref{fig:500_Mgas} and \ref{fig:L500Mgas_L500} which emphasizes the difference to the full-sample fit. As clear from these figures and Table \ref{tab:linearfitgas}, the dispersion of the normal galaxies around the normal galaxy-only fit is reduced relative to that of that seen around the full-galaxy relation. The largest decrease in dispersion appears to be for the longest wavelengths, with the SPIRE500 relation decreasing by a factor of 2, while the dispersion at the shortest wavelengths only decreases slightly.

As the relation for the SPIRE500 and M$_{\rm gas}$ is almost linear for normal galaxies, we can naively assume a 1:1 relation. In this case, we find for the normal KINGFISH galaxies (M$_* > 10^9$\,M$_{\odot}$);
\begin{equation}\label{eqn:MgasL500}
{\rm M}_{\rm gas} [{\rm M}_{\odot}] = 28.5 L_{500} [L_{\odot}] ,
\end{equation}
where $L_{500}$ is the SPIRE 500\mum\ luminosity (i.e.~$\nu L_{\nu}$). Assuming a linear correlation at 500\mum\  only increases the dispersion slightly to 0.118 dex (i.e.~most gas masses within 30\%). 

One issue with the linear relations we present is that they are only monochromatic, and that these rest-frame wavelengths may not be available at higher redshifts when observed using for example Herschel or ALMA. In these cases we suggest a simple linear interpolation of the slope, $B$, and constant, $A$ in Table \ref{tab:linearfitgas}. The slope is monotonic, at least within the range of wavelengths explored here, and so a simple linear interpolation should suffice. For the constant, there appears to be a minimum at 350\mum, however, again a linear interpolation between the bounding wavelengths should be sufficient and lead to only a small uncertainty (as seen by the range of $A$), except at the shortest wavelengths. For continuum measurements outside the range explored by Herschel, an extrapolation based upon a simple modified blackbody can be applied \citep[as done for equation 10 in][]{Scoville14}. 

\subsubsection{The effects of metallicity}

\begin{figure*}
\includegraphics[width=\hsize]{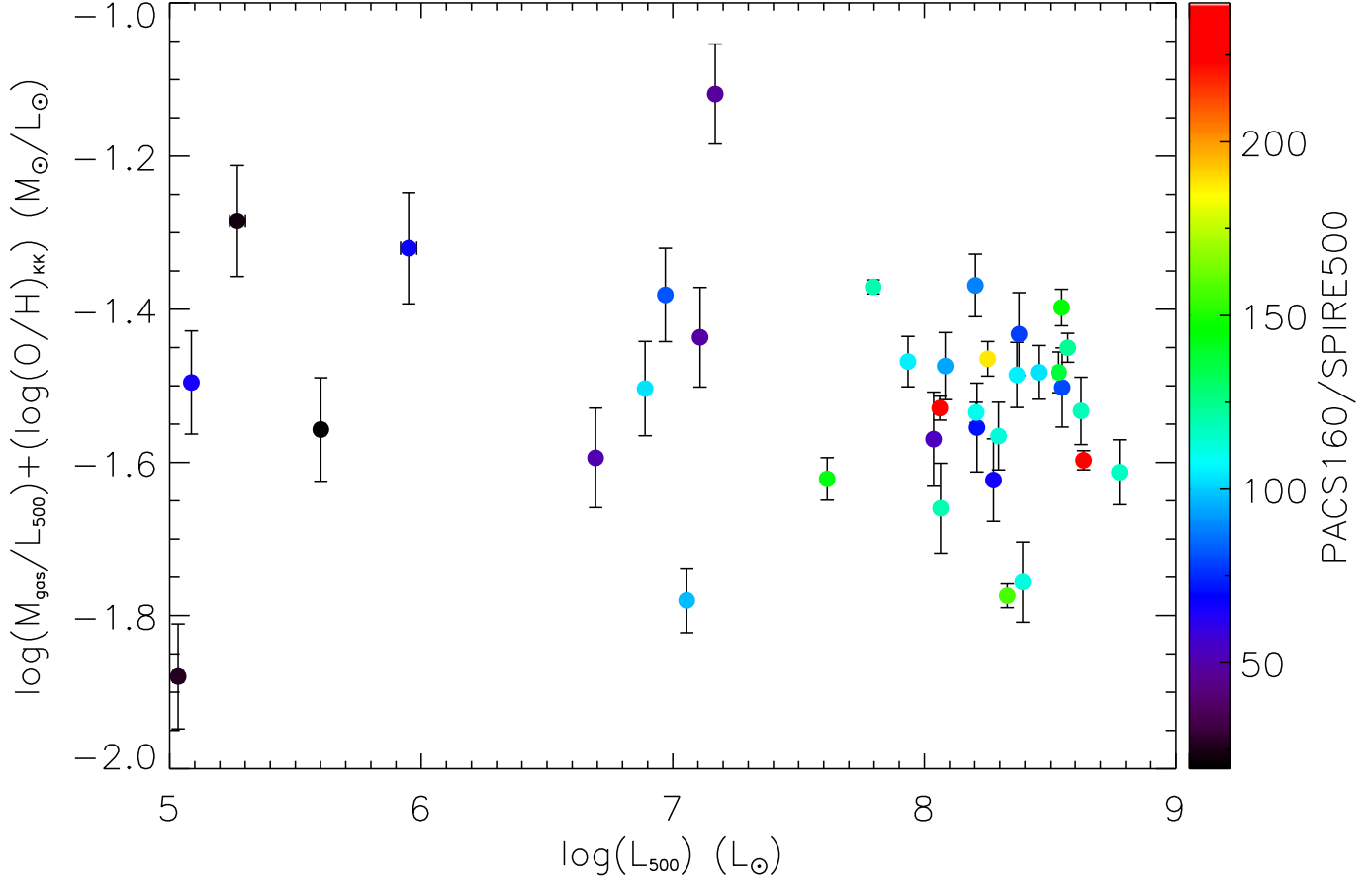}
\caption{SPIRE 500 Luminosity (L$_{500}$) against the ratio of the total gas mass M$_{gas}$ to SPIRE 500 Luminosity as in Figure \ref{fig:L500Mgas_L500}, however corrected for the variation of the dust-to-gas ratio by multiplying each galaxy by the average metallicity determined by \citet{Moustakas10}. This assumes a linear scaling between the metallicity and dust-to-gas ratio. Note that the shown uncertainties do not include that on the metallicity, which can be large and systematic. The color scale shows the 160\mum/500\mum\ flux ratio, a proxy for dust temperature where a lower value indicates warmer dust, and demonstrates that temperature effects do not dominate the observed spread in the corrected ratio.}\label{fig:L500Mgas_corrected}
\end{figure*}

The offset observed at low stellar masses in Figure \ref{fig:L500Mgas_L500}  is clearly associated with a decline in the gas phase metallicity. Given this, the most likely cause for this offset is the variation of the dust-to-gas ratio with metallicity. Naively, we would expect the dust-to-gas ratio (DGR) to at least linearly decrease with metallicity due to the lack of metals from which to form, with recent works suggesting that a linear DGR to metallicity relation is a reasonable assumption \citep[][though the latter work suggests this is true only for $12+$log(O/H)$\gapprox 8$]{Leroy11, Sandstrom13, RemyRuyer14}.
In Figure \ref{fig:L500Mgas_corrected}, we have used the measured metallicity of our galaxies from \citet{Moustakas10} to ``correct'' for a linear variation of the DGR by scaling the M$_{\rm gas}/L_{500}$ ratio by O/H.
With this correction included, the relationship between M$_{\rm gas}/L_{500}$ and $L_{500}$ is basically flat (a fitted slope of $-0.008$), and a dispersion of only $\sigma_{500}=0.147$, much less than the original fits to all galaxies. Thus, over 4 orders of magnitude in IR luminosity, the total gas mass can be determined from the 500\mum\ luminosity within $\sim$40\%, as long as the effects of metallicity on the DGR can be corrected for.

One issue in correcting for the variable DGR is the difficulty in obtaining the gas-phase metallicity of a galaxy at higher redshift. When such information is not available, using  the mass-metallicity relation, presuming a stellar mass for the galaxy has been determined, provides one possible solution \citep[see e.g.,][]{Tremonti04,Kewley08}.

The metallicity of the ISM also affects the relation of dust and gas through the CO luminosity to molecular gas mass conversion factor $\alpha_{\rm CO}$.
For the lowest metallicity galaxies, $\alpha_{\rm CO}$ may increase by an order of magnitude or more from the MW value assumed here because of the presence of ``CO--dark'' molecular gas \citep[see, e.g.][ for the theory and for examples and discussion in the local universe]{Wolfire10, Leroy11}. However, as mentioned in the previous section, the `low-mass galaxies' in our sample all have molecular gas fractions less than 0.01 (given our assumed MW conversion factor, as seen in Figure \ref{fig:fmol_IRratio} in the following section), meaning that the contribution of molecular to the total gas mass in these objects will always be relatively small. This effect will act to increase the offset we observe in Figure \ref{fig:L500Mgas_L500}, but will contribute less than a factor of two (0.3 dex) to any relation. 

\begin{figure*}
\includegraphics[width=\hsize]{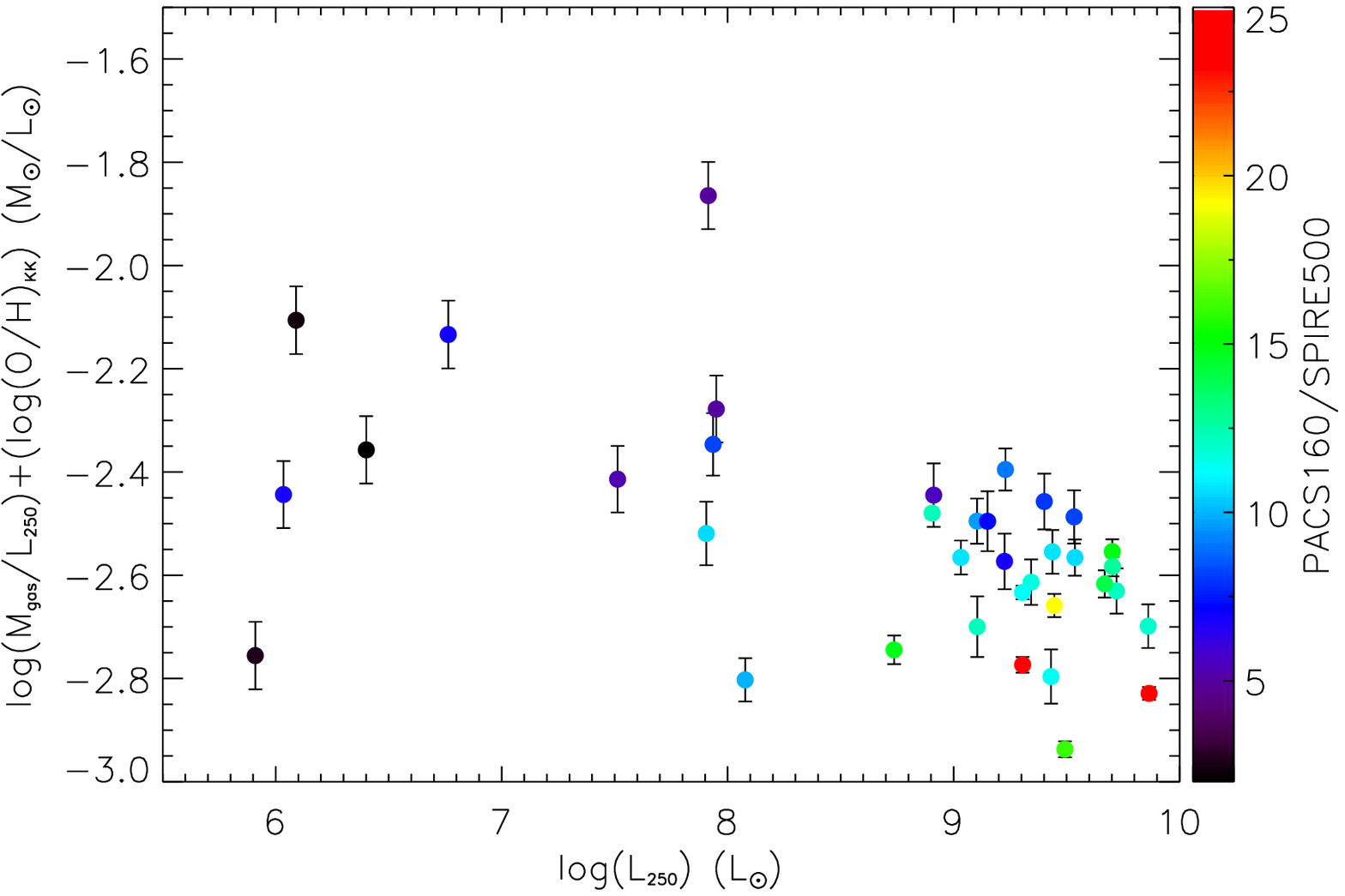}
\caption{Same as Figure \ref{fig:L500Mgas_corrected}, but substituting the SPIRE 250 Luminosity (L$_{250}$) for L$_{500}$. The colors show the PACS160/SPIRE500 color, a measure of the mean dust temperature, where a lower value indicates warmer dust.}\label{fig:L250Mgas_corrected}
\end{figure*}

While the correction for metallicity clearly works for the M$_{\rm gas}-{\rm L}_{500}$ relation at long wavelengths, at shorter wavelengths the effect of dust temperature is still an issue, as suggested by the flattening of the slopes in Table \ref{tab:linearfitgas}.  Figure \ref{fig:L250Mgas_corrected} shows the relation of the SPIRE250 luminosity with M$_{\rm gas}$ corrected for the variable dust-to-gas ratio, as in Figure \ref{fig:L500Mgas_corrected}, with both the negative slope and increased scatter compared to the L$_{500}$ relation are noticeable.
 The color of the points indicates the PACS160/SPIRE500 color, which we use as a proxy for the dust temperature, where a lower value indicates warmer average dust temperatures as the peak of the IR shifts to wavelengths shorter than 160\mum. Note how at high masses (high L$_{250}$), a higher M$_{\rm gas}/{\rm L}_{250}$ is seen with decreasing 160/250 ratio, indicative of warmer temperatures. However, with the M$_{gas}/L_{500}$ plot (Figure  \ref{fig:L500Mgas_corrected}) no such gradient is seen with the PACS160/SPIRE500 ratio, demonstrating the lack of sensitivity of the longer wavelength emission to dust temperature. 
This suggests, not unexpectedly, that to obtain a more accurate gas mass at shorter wavelengths, at least two bands must be observed to correct for dust temperature. However, given that the IR continuum at or around 500\mum\ (600 Ghz) is observable at essentially all redshifts up to $z\sim 6$ using existing ALMA bands (and longer using bands 1 \& 2), and no sensitivity to the IR colors is seen in the scatter at these wavelengths, we suggest using longer rest-frame wavelengths when possible to determine the total gas mass.

\subsection{IR emission versus molecular gas mass}

Given that the star formation rate surface density in nearby galaxies appears to correlate better with the molecular gas surface density than the total gas density \citep[e.g.][]{Schruba11}, determining the molecular gas mass at higher redshift without the necessity of measuring a possibly weak line would be useful.  
When we compare the molecular gas mass (or, more precisely, CO(2-1) luminosity) to the infrared luminosities for our full sample of galaxies we find the wavelengths at the peak of the IR SED (100 and 160\mum) show the tightest correlation, with Table \ref{tab:H2pearson} listing the Pearson correlation coefficient and Spearman rank coefficient for each \emph{Herschel} band for the correlation. We find this is true even when we limit the sample to the normal galaxies. This is somewhat contrary to \citet{Bourne13}, who found the highest correlation of the molecular gas mass with the longest wavelengths. 

\begin{table*}
\begin{center}
\caption{Correlation coefficients for log(M$_{\rm H2}$)\footnote{M81 dwarf B is not detected in CO, and is not included} against individual Herschel band luminosities.}\label{tab:H2pearson}
\begin{tabular}{lrr|rr}
\tableline
 & \multicolumn{2}{c}{All} & \multicolumn{2}{c}{Mass selected}  \\
Band (log$(\nu{\rm L}_{\nu})$& Pearson & Spearman & Pearson & Spearman \\
\tableline
\tableline
  PACS70  &  0.957 &  0.929  &  0.905 &  0.890 \\
  PACS100 &  0.979 &  0.955 &  0.946 &  0.925 \\
  PACS160 &  0.977 &  0.969 &  0.959 &  0.949 \\
  SPIRE250 &  0.980 &  0.958 &  0.951 &  0.932  \\
  SPIRE350 &  0.974 &  0.936 &  0.933 &  0.895  \\
  SPIRE500 &  0.964 &  0.901 &  0.908 &  0.836  \\
  \tableline
 \end{tabular}
\end{center}
\end{table*}

\begin{figure*}
\includegraphics[width=\hsize]{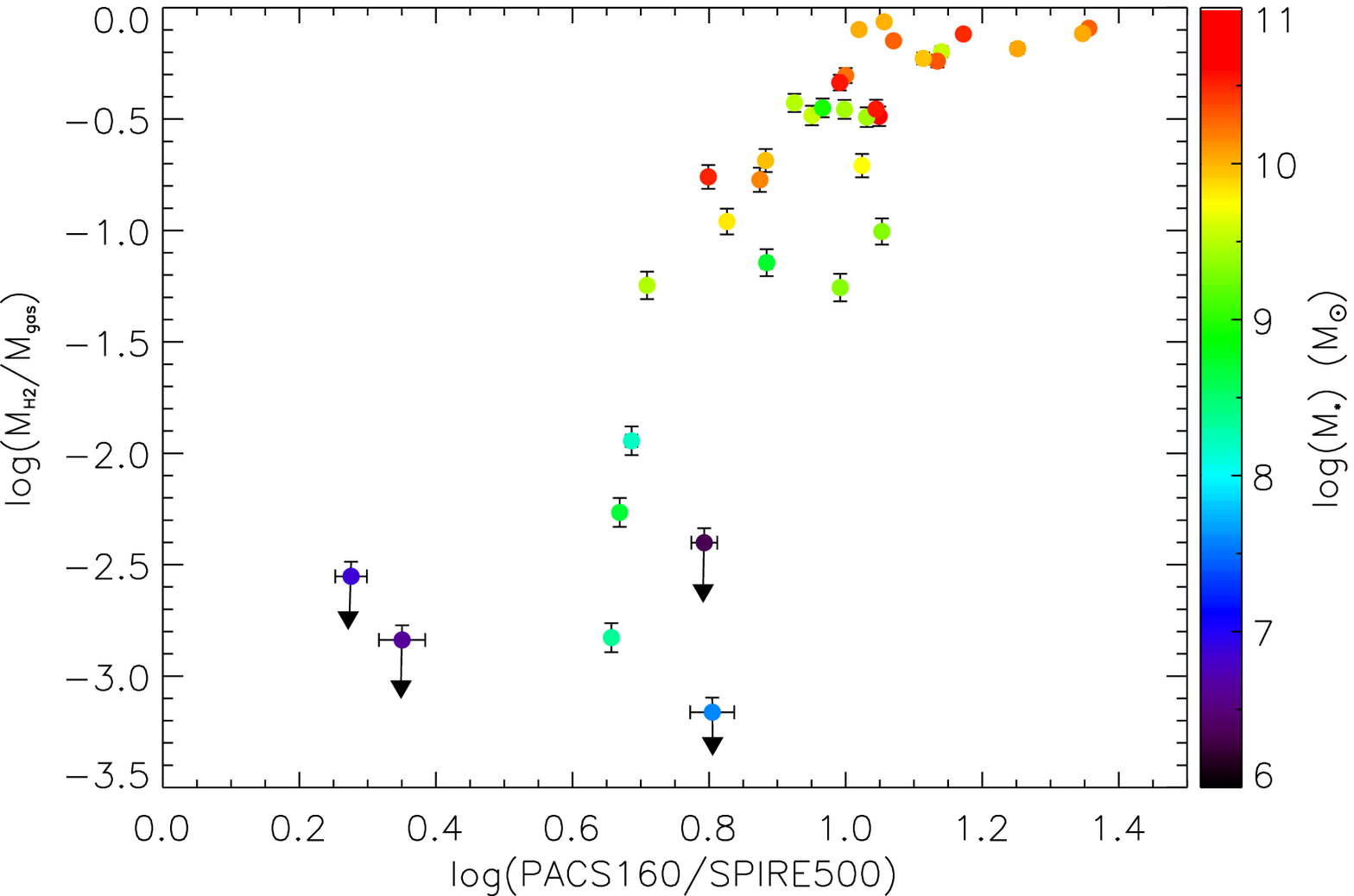}
\caption{The molecular gas fraction (M$_{\rm H2}/{\rm M_{gas}}$) as a function of the PACS160/SPIRE500 flux ratio, a proxy for mean dust temperature (low values indicate warmer dust).  The color scale indicates the total stellar mass of the galaxies.}\label{fig:fmol_IRratio}
\end{figure*}

This is surprising given that molecular gas would be expected to be cold, and therefore mostly associated with cold dust. However a closer examination reveals that this is exactly the case, as demonstrated in Figure \ref{fig:fmol_IRratio}. In this figure we plot the molecular gas mass fraction, M$_{\rm H2}/{\rm M_{gas}}$, as a function of the dust temperature as represented by the PACS160/SPIRE500 flux ratio (a higher ratio indicates colder dust as the IR peak shifts to 160\mum). Given that the denominators correlate, as shown in Figure \ref{fig:500_Mgas}, this removes any mass scaling, and acts to highlight the correlation between the 160\mum\ luminosity and molecular gas mass indicated in Table \ref{tab:H2pearson}. This figure reveals that as the dust gets colder (higher 160\mum/500\mum) the molecular gas fraction increases, with both these quantities correlated with the galaxy stellar mass (as shown by the color scale in Figure \ref{fig:fmol_IRratio}), or similarly the average gas metallicity. 

The evolving shape of the IR SED with the metallicity of the galaxy can be seen clearly in Figure 8 in \citet{RemyRuyer13}, demonstrating the warmer average dust temperatures in the lower metallicity objects. The warmer temperatures in these galaxies are likely caused by the harder and stronger interstellar radiation fields (ISRF). The ISRF is harder due to the lower stellar metallicity in these galaxies, while the ISRF tends to be stronger due to the lower dust density (given the lower dust to gas ratio) leading to decreased dust shielding and a longer mean free path of UV photons \citep{Madden06}.

This lower dust shielding will lead to a lower molecular gas fraction due to increased H$_2$ dissociation, and a relatively higher dissociation of the CO molecule, and thus an increased $\alpha_{\rm CO}$ factor. As we assume a constant $\alpha_{\rm CO}$ factor,  an increasing $\alpha_{\rm CO}$ with metallicity \citep[due to the presence of ``CO--dark'' molecular gas,][]{Wolfire10}, will also cause the decrease in the molecular gas fraction that we observe \citep[if a variable $\alpha_{\rm CO}$ is assumed, it is uncertain how much the molecular gas fraction will change with metallicity, see e.g.][specifically their Figure 3]{RemyRuyer14}.

Once the correlation of these ratios with stellar mass is accounted for, no association with other galaxy parameters such as SFR or sSFR is seen \citep[note however that these parameters correlate with stellar mass in our galaxy sample, as seen in SDSS galaxies, e.g.][]{Brinchmann04}. This is surprising, given that the 160/500 ratio traces dust temperature, and that molecular gas is associated with star formation. However, it may be that smaller scale effects not seen in the integrated quantities, such as the spatial distribution of gas, dust and stars, is affecting the correlation to an extent that 2nd order effects are not seen. 

In general, we find the best power-law fit to be:
\begin{equation}
\log {\rm M_{H2}} [ {\rm M_{\odot}}] = -2.86 + 1.22 \log (\nu{\rm L}_{\nu}(160\mu m)/{\rm L}_{\odot})
\end{equation}
 with a dispersion of $\sigma=0.32$ around this relation. However once a galaxy is molecular rich, it is uncertain how well the 160\mum\ luminosity will trace M$_{\rm H2}$. As can be seen in Figure \ref{fig:fmol_IRratio}, at high molecular fractions a broad range of 160\mum/500\mum\ colors are possible. Thus for the most massive objects, the 500\mum\ luminosity is a better tracer of the molecular gas simply because M$_{H2}/M_{\rm gas}\sim 1$. Similarly, given the uncertain effects of dust heating on this relation, extrapolating this to higher redshifts may be problematic.

\subsection{Extension to higher luminosities and redshifts}

The sample of galaxies we explore here are typical local galaxies, and cover a range of luminosities and stellar masses (as seen in Table \ref{tab:galaxies}), however these galaxies may not be representative of more luminous IR galaxies typically seen in sub-mm surveys. 
\citet{Bourne13} explore a slightly higher mass range than our sample, though with a significant overlap. For our full sample we find much flatter slopes than \citet{Bourne13} found for the IR versus HI, CO(2-1) and CO(3-2) luminosities. However for our more massive ``normal'' galaxy sample that overlaps the range observed in \citet{Bourne13}, our slopes appear very similar to those determined by \citet[][specifically their Figure 4]{Bourne13}, at least for HI and CO(2-1), the lines we use for our total gas mass. However, while they do see trends with stellar mass and metallicity (such as the decrease in L$_{\rm CO}$/M$_{\rm HI}$ with mass and metallicity), their dynamic range is not large enough to determine the relations we see here. 

\citet{Scoville14} explored the correlation of sub-mm luminosity with ISM mass in a sample of sub-mm galaxies at $z = 2 -3$, in addition to a local sample that overlaps our sample with 850\mum\ detections, and a Planck observation of our Galaxy. Assuming that all the gas was molecular and a similar conversion factor ($\alpha_{\rm CO}=4.6$), \citet{Scoville14} find a ratio of L$_{850}$/M$_{\rm gas}=\alpha_{850}= 1\pm0.5\times10^{20}$ erg\,s$^{-1}$\,Hz$^{-1}$\,M$_{\odot}^{-1}$ for the sub-mm galaxies, and similar values for the local galaxies (for which the HI mass was also included). Rearranging Equation \ref{eqn:MgasL500}, we find a ratio at 500\mum\ of $\alpha_{500} =2.26\times10^{20}$ erg\,s$^{-1}$\,Hz$^{-1}$\,M$_{\odot}^{-1}$ for our normal galaxy sample, which, assuming a dust emissivity with a power-law of 1.8, corresponds to $\alpha_{850}=0.87$ erg\,s$^{-1}$\,Hz$^{-1}$\,M$_{\odot}^{-1}$, in good agreement with the sub-mm galaxy sample of \citet{Scoville14}. 

However, applying a single sub-mm emission to gas mass conversion factor, $/alpha$, to all galaxies removes any dependence on other galaxy properties. We find that for normal galaxies in the local Universe, an assumption of a linear relation between sub-mm luminosity and gas mass is reasonable. Given the range of IR luminosities explored in our work and in \citet{Scoville14}, it suggests that, while dust temperatures will affect the sub-mm luminosity to gas mass relationship, the variation in temperature is small enough that it has only a small impact in the Rayleigh-Jeans regime and thus the relationship.
Conversely, there appears to be a significant metallicity evolution over cosmic time \citep[see e.g.][]{Yabe14}. Therefore to apply the observed L$_{\rm sub-mm}$--M$_{\rm gas}$ relation to all galaxies at higher redshifts, it is crucial that this evolution of the metallicity be accounted for.  This can be done through a direct correction for an inferred metallicity, either directly measured or determined from a mass-metallicity relation, or through allowing the relation (or $\alpha$) to evolve with redshift using the metallicity evolution as a basis for the evolving DGR. 

An evolving DGR was allowed for in the work of \citet{Magdis11,Magdis12}, who calculated the total gas mass from IR observations in a sample of ``normal'' high-redshift galaxies ($z\sim 1-4$). Instead of using a direct correlation of sub-mm luminosity and gas mass, they first calculated the dust mass using model fits to their observed IR SEDs \citep[such as the][model]{Draine07a}, and then converted to the total gas content using an empirical relationship between the DGR and metallicity determined locally \citep[][]{Leroy11}.  While this two-step approach may introduce systematic errors, \citet{Magdis12} find their determined gas masses consistent with the CO luminosities (assuming Milky Way conversion factors that also vary with metallicity, and assuming the gas is mostly molecular). 

An example of the application of using IR fluxes to determine the evolution of gas mass is the work of \citet{Santini14}, who stacked the IR fluxes of galaxies with similar stellar masses, star formation rates and redshifts, to create representative IR SEDs. They then fit these SEDs using dust models, including simple modified blackbodies, to determine the dust masses of these objects, which they converted to a gas mass allowing for a variable dust-to-gas ratio dependent upon stellar mass and SFR.  Using these they find a strong correlation in the gas fractions of galaxies with their stellar mass and SFRs, and find strong evolution in both the gas fraction and star formation efficiency (SFR/M$_{\rm gas}$) with redshift.

\section{Radial trends and galaxy sizes}\label{sec:radial}

In the previous section, we only considered the galaxies as integrated objects, with a single gas mass and luminosity. However, the KINGFISH sample provides much more detail than this, and the sub-arcsecond capabilities of ALMA will mean, even at high-redshift, the continuum emission from galaxies can potentially be resolved \citep[see, e.g.][]{Hodge13}.
Thus to explore the gas -- L$_{500}$ relation further, we break the galaxies into elliptical annuli using the position angles and axis ratios of the galaxies given in Table \ref{tab:galaxies}. The annuli are chosen to be in steps of 2\,kpc along the major axis

\begin{figure*}
\includegraphics[width=\hsize]{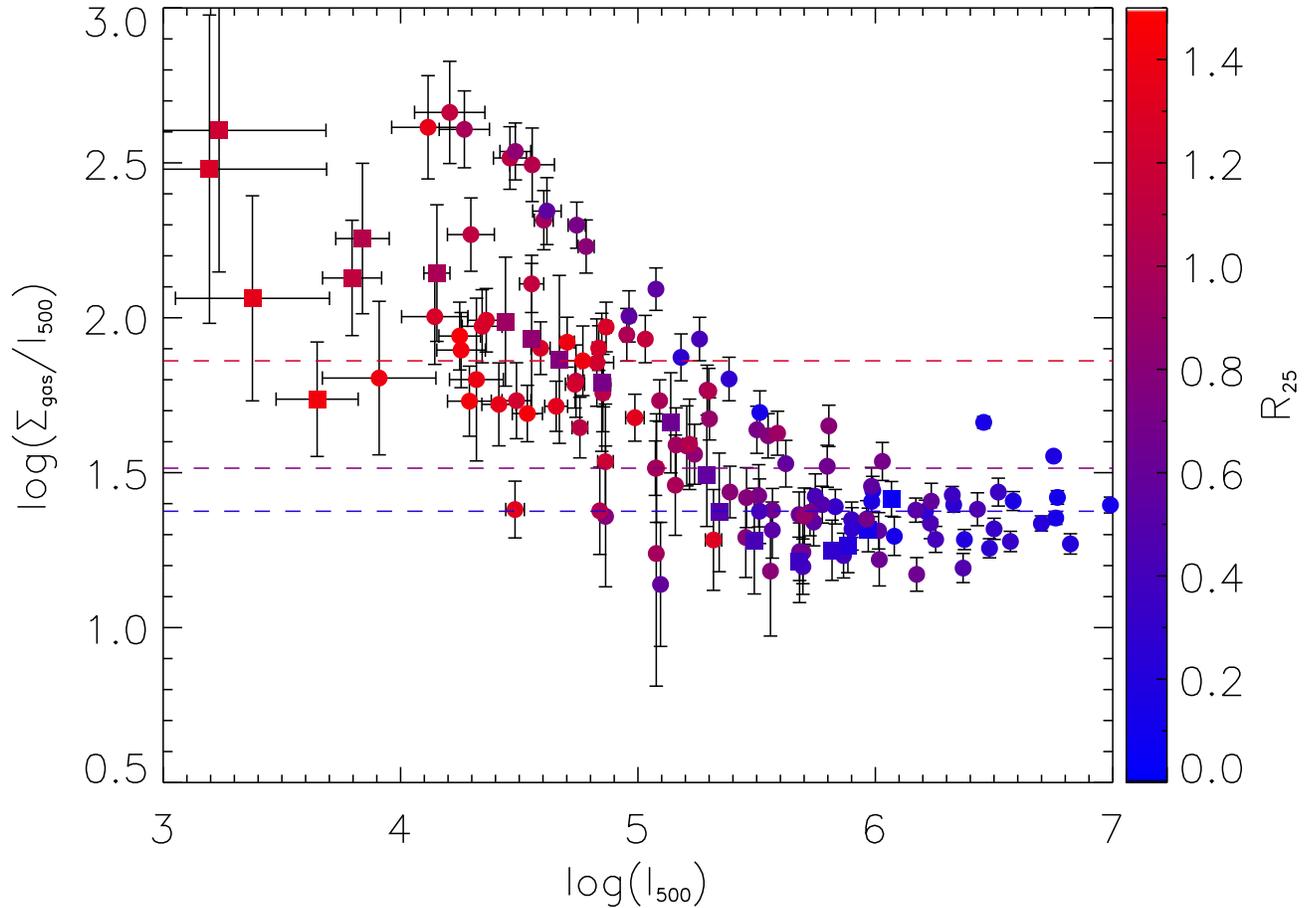}
\caption{SPIRE 500\mum\ luminosity (L$_{500}$) against the ratio of the total gas mass M$_{gas}$ to L$_{500}$ (following Figure \ref{fig:L500Mgas_L500}), measured in elliptical annuli of $\Delta r=2$\,kpc for galaxies in our sample. The colors indicate the position of the annuli in terms of the optical radius $R_{25}$ of each galaxy, as given in Table \ref{tab:galaxies}. The dashed horizontal lines indicate the median S$_{gas}$/I$_{500}$ ratio for annuli between $0-0.5$, $0.5-1.0$, and $1.0-1.5\, R_{25}$ as indicated by the colors..}\label{fig:annuliSgasI500}
\end{figure*}

In Figure \ref{fig:annuliSgasI500} we plot the ratio of the gas mass surface density and SPIRE500 surface brightness for all annuli in our galaxy sample that satisfy the criteria that the annuli was larger than the resolution of the SPIRE500\mum\ beam (36\arcsec). The colors of the data indicate the position of the annuli in terms of the optical radius $R_{25}$ of each galaxy, as given in Table \ref{tab:galaxies}. This figure demonstrates that within galaxies the radial profiles follow the same pattern as seen for the integrated luminosities (i.e.~as in Figure \ref{fig:L500Mgas_L500}), albeit with a larger scatter. Some of this scatter arises from faint background sources at large radii which have not been masked. As with the luminous galaxies, we see that for high surface brightness regions $\Sigma_{gas}$/I$_{500}$ is linear, but when we move to lower I$_{500}$ we find an increasing trend of $\Sigma_{gas}$/I$_{500}$. This trend is affected by both the intrinsic metallicities of the galaxies as well as the radial trends within galaxies, with a higher $\Sigma_{gas}$/I$_{500}$ ratio observed at larger radii. This is emphasized by the horizontal lines which indicate the median $\Sigma_{gas}$/I$_{500}$ ratio for annuli between $0-0.5$ (blue), $0.5-1.0$ (purple), and $1.0-1.5\, R_{25}$ (red). This radial $\Sigma_{gas}$/I$_{500}$ gradient is likely associated with the radial metallicity gradient within each galaxy, and when the observations are corrected for the metallicity gradient determined by \citet{Moustakas10} for some of our sample, this gradient is reduced. However the gradient is not totally removed, and has a large scatter introduced by the uncertainties in the metallicity gradient \citep[see][for a discussion on these uncertainties]{Moustakas10}.

\begin{figure*}
\includegraphics[width=\hsize]{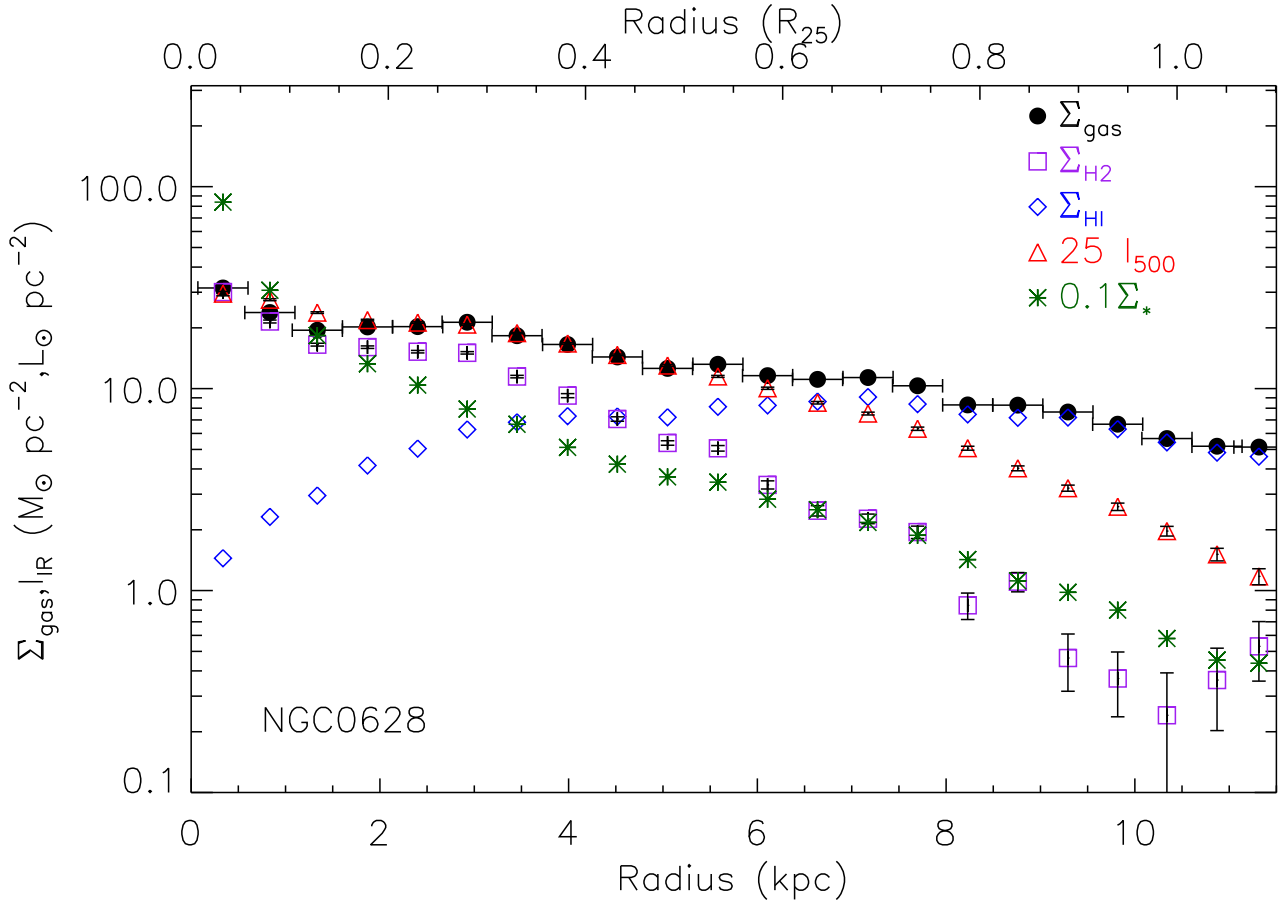}
\caption{Radial variation of the SPIRE 500 surface brightness, $I_{\rm IR}$ (triangles), and the HI (diamonds), H$_2$ (squares), and total gas (solid circles) and stellar mass (stars) surface densities in bins of 15\arcsec\ in NGC\,628. The SPIRE 500 surface brightness and stellar mass surface density have been scaled by 25 and 0.1, respectively, to place these on the same scale as the gas mass surface densities. The HI, H$_2$, and total gas mass surface densities are from \citet{Schruba11}, and the stellar mass surface density is from Querejeta et al., (in prep.).} \label{fig:ngc628radial}
\end{figure*}

Yet using the longer wavelength continuum emission is still a robust tracer of the total gas mass of galaxies, and the gas surface density for most of a galaxy. The robustness of the sub-mm continuum can be seen in Figure \ref{fig:ngc628radial}, where we compare the radial variation of the SPIRE500 mean surface brightness, $I_{\rm 500}$, against the mean gas mass surface density, $\Sigma_{\rm gas}$, in NGC\,628. Using the radial determinations of the HI and H$_2$ gas mass surface densities from \citet{Schruba11}, who stacked the HERACLES CO observations to detect the low column density molecular gas at large radii, we can determine both the radial variation of the total gas mass and the molecular gas fraction in our sample. Using the same 15\arcsec\ elliptical annuli as \citet{Schruba11}, we have determined the radial variation of the SPIRE500 surface brightness (note that 15\arcsec\ under samples the 36\arcsec\ SPIRE 500 beam by a factor of 3). For illustration purposes, we have scaled $I_{\rm 500}$ by a factor of 25 to be on the same scale as the total gas mass surface density. It is clear that the IR traces better the total gas mass surface density instead of either the H$_2$ or HI gas alone in the inner radii.  However, at radii beyond 0.7 $R_{25}$ there is an increasing deviation of the IR brightness relative to the gas mass density. A similar trend is seen for some of the other galaxies in our matched sample, such as NGC\,925. However the trend is not seen in all galaxies, with resolution and inclination affecting the observed trends in some objects (see the attached figures in the appendix). 

The point where the sub-mm surface brightness and gas mass density separate does not appear to directly correlate with the radius at which the gas becomes mostly atomic, emphasizing that the dust is not just tracing the molecular gas. This conclusion is further supported by galaxies which are predominantly atomic at all radii. 
Neither does this separation appear to be related to the surface mass density of NGC\,628 becoming gas dominated. In Figure \ref{fig:ngc628radial} we also indicate the stellar surface density determined from Spitzer 3.6\mum\ data by Querejeta et al. (in prep.) using the data from the S4G survey \citep{Sheth10}, arbitrarily scaled by 0.1 to be on the same scale as the gas mass density. The disk of NGC\,628 becomes gas dominated beyond the radius at which the IR deviates from $\Sigma_{\rm gas}$ (at $R_{25}\sim1$). 

\begin{figure*}
\includegraphics[width=\hsize]{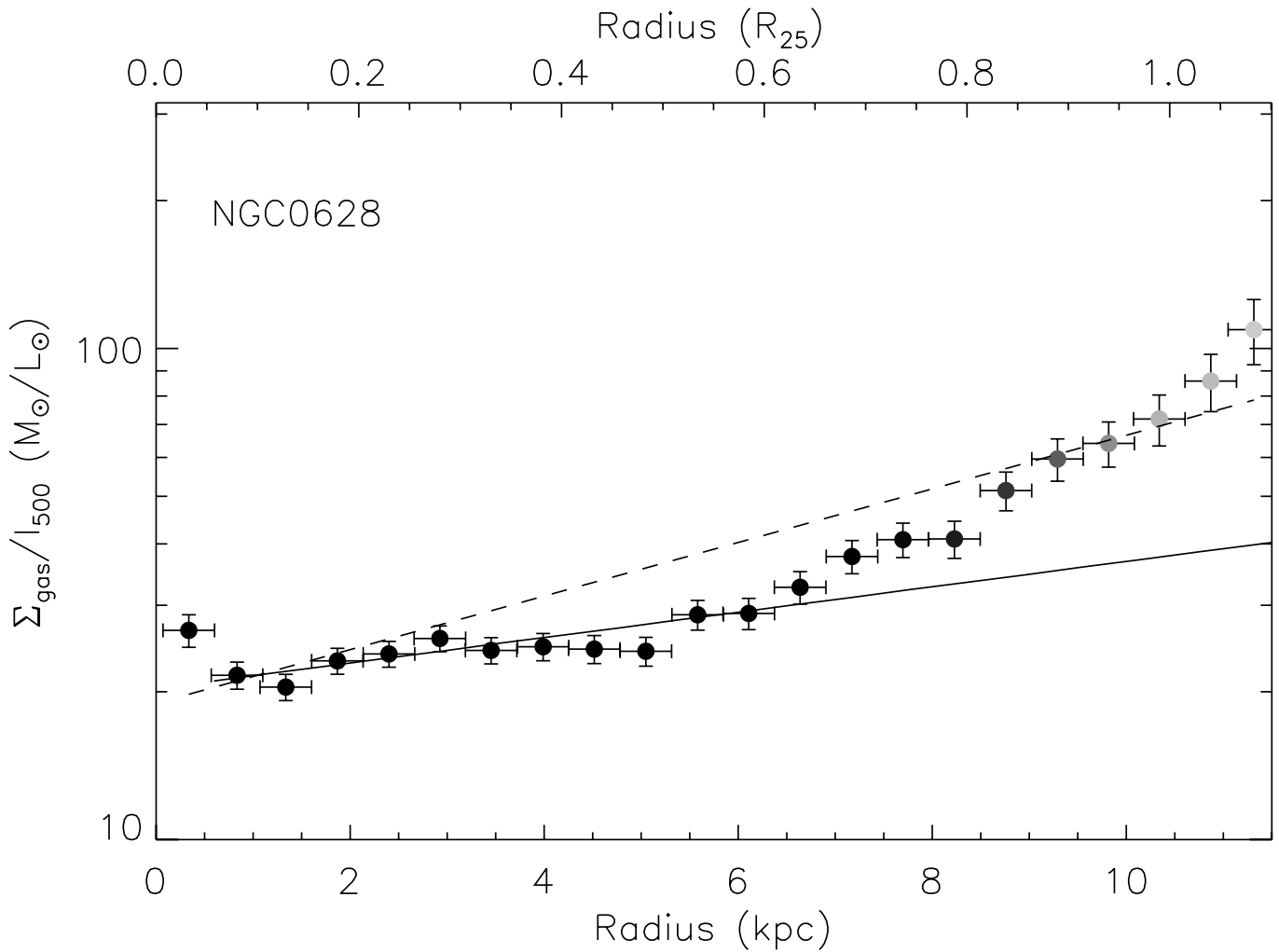}
\caption{Radial variation of the ratio of gas mass surface density to SPIRE 500 surface brightness, $\Sigma_{\rm gas}/I_{\rm 500}$ for the face on spiral NGC\,628. At $\sim 0.7 R_{25}$, a change in how the ratio varies with radius is visible. The greyscale shading of the data indicate the fraction of significant pixels with signal-to-noise greater than 3. Overplotted  are two lines showing the inverse of the metallicity gradient determined by \citet{Moustakas10} using the \citet{Pilyugin05} (solid line) and \citet{Kobulnicky04} (dashed line) calibrations. Both lines have been normalized to the value of $\Sigma_{\rm gas}/I_{\rm 500}$ at $0.1 R_{25}$.
} \label{fig:ngc628gdr}
\end{figure*}

The deviation of the sub-mm emission from the gas is more clearly seen in Figure \ref{fig:ngc628gdr}, where we plot the ratio of mean gas mass surface density to 500\mum\ surface brightness as a function of radius for NGC\,628. At the inner radii of the galaxy, the $\Sigma_{\rm gas}/I_{500}$ ratio is flat and then increases in slope beyond $R_{25}\sim0.7$. To demonstrate that this change is significant, we have shaded the data to indicate the fraction of pixels that are significant ($S/N>3$), with darker points indicating a larger fraction. This shading is conservative as we are integrating in annuli, meaning that we are more sensitive to diffuse emission at larger radii. Even given this conservative estimate, it is clear that the measured values of $\Sigma_{\rm gas}/I_{500}$ ratio are significant out to $R_{25}\sim1$.

One possibility for this deviation is that we are seeing a change in the dust to gas ratio at these radii, possibly related to a variation in the radial metallicity gradient at the larger radii. However, \citet{Moustakas10} measure gas-phase abundances out to $\sim R_{25}$ for several of our galaxies, including NGC\,628 (see their Figure 7), and no such changes in the abundance gradients are seen.  The expected variation of the gas to dust ratio is indicated by the inverse of the metallicity gradient determined by \citet{Moustakas10},  using both the \citet{Kobulnicky04} or \citet{Pilyugin05} calibrations, as shown by the dashed and solid lines, respectively, on Figure \ref{fig:ngc628gdr}. These lines are both normalized to the value of $\Sigma_{\rm gas}/I_{500}$ at $0.1 R_{25}$ to emphasize the similarity (or difference) of the determined metallicity slopes compared to the observed variation of the $\Sigma_{\rm gas}/I_{500}$ ratio. However, the lines also demonstrate the issues in interpreting the metallicity gradient. The \citet{Pilyugin05} calibration follows reasonably well the observed $\Sigma_{\rm gas}/I_{500}$ ratio up to $R_{25}\sim0.7$, beyond which a deviation from this slope is clearly visible. Yet the \citet{Kobulnicky04} calibration shows a much steeper slope, and, while reasonable below $0.4 R_{25}$, over-estimates the observed ratio between 0.4 and 0.8 $R_{25}$, and no deviation of the $\Sigma_{\rm gas}/I_{500}$ ratio relative to this gradient is seen at larger radii. Which gradient is more representative of the true gas-phase abundance gradient is uncertain, though \citet{Croxall13} find a metallicity in between the two calibrations shown here at $R_{25}\sim0.4$ using temperature insensitive metallicity calibration based on FIR fine structure lines, suggesting that the gradient may lay between these lines.

The $\Sigma_{\rm gas}/I_{500}$ radial variations for the other galaxies in our sample are included in the online-only appendix. For all galaxies this ratio lies between $20-30$, with the ratio becoming larger at larger radii. Several galaxies also show a turn up in $\Sigma_{\rm gas}/I_{500}$ within $0.1R_{25}$, such as NGC\,6946. The increase in the $\Sigma_{\rm gas}/I_{500}$ in these galaxy centres is likely associated with the decreased central $\alpha_{\rm CO}$ values observed by \citet{Sandstrom13} in several of our galaxies, including NGC\,6946. A central decrease in $\alpha_{\rm CO}$ would mean that we overestimate $\Sigma_{\rm gas}$ at the center of our galaxies due to our assumption of a constant Milky-Way value, and hence over-estimate the $\Sigma_{\rm gas}/I_{500}$ ratio. 
For galaxies for which it is available, we have also plotted the inverse of the determined metallicity gradients as in Figure \ref{fig:ngc628gdr}. For several galaxies (e.g.~NGC\,3198) the metallicity gradient is sufficient to explain the observed radial trend in $\Sigma_{\rm gas}/I_{500}$, with no significant deviations seen. For some galaxies, i.e.~NGC\,4254 and NGC\,4321, the metallicity gradients are much steeper than the observed $\Sigma_{\rm gas}/I_{500}$ ratio. Given the observed gradients in gas and stellar surface mass densities, it is possible that the metallicity gradients are much flatter than have been determined for these objects. Finally for several galaxies the same excess at large radii relative to the metallicity gradient as in NGC\,628 is seen (e.g.~NGC\,925 and NGC\,6946).

For these galaxies which show a radial deviation in the $\Sigma_{\rm gas}/I_{500}$ ratio, like NGC\,628 and NGC\,925, another possibile explanation is that the different conditions in the outskirts of the galaxies lead to different dust properties which result in different emissivities. Such a variation in the emissivity with radius for some of the KINGFISH galaxies has already been suggested by \citet{Galametz14} using sub-mm data and in M33 by \citet{Tabatabaei14}.

\begin{figure*}
\includegraphics[width=\hsize]{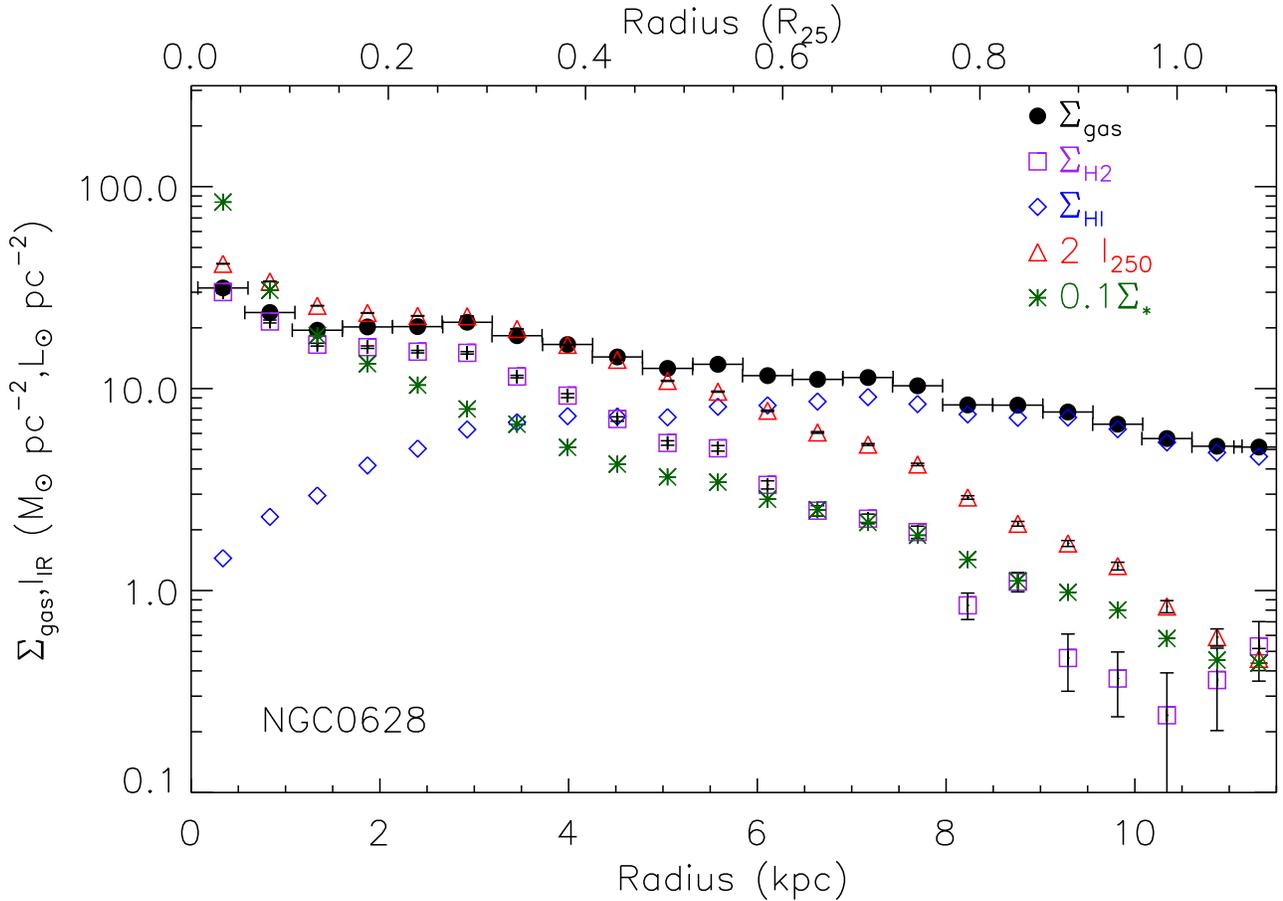}
\caption{Same as Figure \ref{fig:ngc628radial} but with the radial variation of the SPIRE 250 surface brightness in NGC\,628. The SPIRE 250 surface brightness has been scaled by only 2 to place it on the same scale as the mass surface densities as it is closer to the peak of the IR SED. The HI, H$_2$, and total gas mass surface densities are from \citet{Schruba11}.} \label{fig:ngc628radial_S250}
\end{figure*}

However, at least some part of this deviation at large radii must arise from the decrease in the mean interstellar radiation field available to heat the dust at these radii, as demonstrated by the radial decrease in stellar density relative to gas density in Figure \ref{fig:ngc628radial}. A weaker interstellar radiation field will lead to lower average dust temperatures and thus correspondingly weaker IR emission. This can be seen in Figure \ref{fig:ngc628radial_S250}, where we plot the same quantities as in Figure \ref{fig:ngc628radial} for NGC\,628, but now with the shorter wavelength SPIRE250\mum\ band (note the different scaling factor). While it is clear in this figure that a change in the ratio of gas mass to IR luminosity occurs at a similar radius, the ratio decreases much more rapidly.  This implies a change in the IR slope, and thus either a decrease in dust temperature or a flattening of the dust power-law emissivity. A decreasing dust temperature would lead to a lower IR luminosity even without a variation of the dust to gas ratio. 

In any case, it is clear from Figure \ref{fig:ngc628radial} and the other galaxies in the Appendix that for the outskirts of many galaxies (beyond $\sim 0.7 {\rm R}_{25}$), the IR emission no longer traces the total gas mass in the same manner. For the galaxies in our sample, the total gas mass is dominated by the atomic gas at these radii. For many of our galaxies a significant fraction of HI is found outside the optical radius (a median of 40\% for our sample), and for several objects emission can be found even outside 2$R_{25}$, with an obvious (and unrepresentative) example being NGC3077 with its extended tidal streams \citep{Walter11}.  By defining our apertures purely on the extent of the optical and IR emission, it is obvious that such extended gas disks will not be accounted for in the above relations, with the relations missing a large fraction of the total gas mass in some galaxies, like the HI rich dwarfs like DDO154.  However it is still an open question how much relevance this extended HI gas has on the galaxy as a whole, with all of the molecular gas and most of the star formation inside our chosen ``total'' galaxy apertures and it being uncertain whether this gas external to the galaxy will participate at all in future star formation events within the galaxy.  

\section{Summary}\label{sec:summary}
The evolution of the gas mass density over cosmic time plays a direct role in the evolution of galaxies through its link to the star formation rate. Yet measuring the gas mass directly of a large number of galaxies at high redshifts through line observations is technically difficult, and time consuming. As dust and gas are intimately associated, the dust infrared continuum may provide a more feasible way to determine the total gas mass of galaxies at high redshift. To demonstrate the use of the IR continuum as a gas tracer we have compared the total gas mass and IR luminosity of a well-studied sample of 36 nearby galaxies (average distance~10\,Mpc) observed with the KINGFISH, THINGS, and HERACLES surveys. These galaxies sample the peak of the mass function, with stellar masses from $\sim10^{6.5}$ to $\sim10^{10.5}\,  {\rm M}_{\odot}$, and, unlike the local merger-induced ULIRGs, these galaxies have disk-like morphologies that are likely more representative of the main-sequence galaxies at high redshift observed in current and future deep sub-mm surveys.

We find a strong correlation between the total gas mass and IR luminosity in all Herschel bands, with the strongest and tightest correlation found for the longest wavelength (SPIRE 500\mum). The gas to luminosity ratio, ${\rm M}_{\rm gas}/{\rm L}_{500}$, is found to increase with decreasing sub-mm luminosity, which can be ascribed to the declining metallicity associated with the lower galaxy masses of the low luminosity galaxies. This declining metallicity will lead to a lower dust to gas ratio, and therefore higher gas to dust luminosity ratio. 
We provide fits to the total gas mass versus IR luminosity relations for all Herschel bands, which can be interpolated between to obtain an empirical relation to be used at any IR wavelength for determining the total gas mass (whose coefficients are given in Table \ref{tab:linearfitgas}).
To minimize the effects of metallicity, we also fit only galaxies with stellar masses greater than $10^9\,{\rm M}_{\odot}$ in our sample, which also results in a significantly reduced dispersion. For the SPIRE 500 band we find that a linear relation is a reasonable approximation, and find that log$({\rm M}_{\rm gas}/{\rm L}_{500})=28.5\, [{\rm M}_{\odot}/{\rm L}_{\odot}]$, with a dispersion of only 0.118 dex.

The molecular gas mass surface density has been demonstrated to be more tightly associated with the SFR than the atomic gas, thus we also provide calibrations for the molecular gas. We find similarly strong correlations, with the strongest correlation found for the PACS100 and PACS160 bands that lie at the peak of the IR emission, and are strongly correlated with the total IR luminosity. This correlation between L$_{160}$ and M$_{\rm H2}$ appears to be driven mostly by the correlation of both the average dust temperature (as measured by PACS160/SPIRE500) and molecular gas fraction (L$_{\rm CO}$/M$_{\rm gas}$) with galaxy mass and metallicity, with dwarf galaxies showing both warmer dust and lower molecular gas fractions. Given this correlation, we suggest that the L$_{160}$--M$_{\rm H2}$ relation not be applied to higher redshifts.

Using the resolved nature of the IR and gas surveys we employed in this survey, we also explored the correlations using elliptical annuli, and find the same correlations as the integrated comparisons, albeit with larger scatter. For most galaxies we find that the $\Sigma_{\rm gas}/I_{500}$ ratio increases with radius, as expected from the determined radial metallicity gradients. However, for some galaxies the $\Sigma_{\rm gas}/I_{500}$ relation is observed to deviate from the metallicity gradient at larger radii, becoming much larger.  A drop in the mean dust temperature plays at least some role in this observed excess, given the decrease in stellar mass surface densities relative to gas surface densities, and supported by the observed steeper $\Sigma_{\rm gas}$ to $I_{250}$ relation relative to the $I_{500}$ radial relation. This deviation in the $\Sigma_{\rm gas}/I_{500}$ ratio in the outskirts in galaxies, and the extended HI disks not visible in the IR observed in a fraction of our galaxies, suggest that the mass of extended gas in galaxies cannot be straightforwardly determined from the IR emission.

However, even with these radial trends, the total gas mass of galaxies within their optical radius can be well determined within a factor of 2 by observations of the sub-mm dust emission, and typically to within 30\% for typical (log(M$_*/{\rm M}_{\odot})>9$) galaxies observed in high redshift samples.

\acknowledgements
The authors would like to thank the anonymous referee whose comments helped expand and improve the details of this work.



\end{document}